\input lanlmac
\def\href#1#2{{#2}}

\input epsf.tex

\overfullrule=0mm

\newcount\figno
\figno=0
\def\fig#1#2#3{
\par\begingroup\parindent=0pt\leftskip=1cm\rightskip=1cm\parindent=0pt
\baselineskip=11pt
\global\advance\figno by 1
\midinsert
\epsfxsize=#3
\centerline{\epsfbox{#2}}
\vskip 12pt
{\bf Fig.\ \the\figno:} #1\par
\endinsert\endgroup\par
}
\def\figlabel#1{\xdef#1{\the\figno}}
\def\encadremath#1{\vbox{\hrule\hbox{\vrule\kern8pt\vbox{\kern8pt
\hbox{$\displaystyle #1$}\kern8pt}
\kern8pt\vrule}\hrule}}


\def\IR{\relax{\rm I\kern-.18em R}}
\font\cmss=cmss10 \font\cmsss=cmss10 at 7pt

\font\cmss=cmss10 \font\cmsss=cmss10 at 7pt
\def\IZ{\relax\ifmmode\mathchoice
{\hbox{\cmss Z\kern-.4em Z}}{\hbox{\cmss Z\kern-.4em Z}}
{\lower.9pt\hbox{\cmsss Z\kern-.4em Z}}
{\lower1.2pt\hbox{\cmsss Z\kern-.4em Z}}\else{\cmss Z\kern-.4em Z}\fi}
\def\IN{\relax{\rm I\kern-.18em N}}
\def\circbullet{{\bigcirc \kern-.75em \bullet \kern .3em}}
\def\circcirc{{\bigcirc \kern-.75em \circ \kern.3em}}
\def\smallcircbullet{{{\scriptscriptstyle{\bigcirc \kern-.5em \bullet}} \kern .2em}}
\def\smallcirccirc{\relax{{\scriptscriptstyle{\bigcirc \kern-.5em \circ}} \kern .2em}}
\def\b{\circ}
\def\n{\bullet}

\def\gbbbb{\Gamma_4^{\hbox{$\scriptstyle \b \b$}\kern -8.2pt
\raise -4pt \hbox{$\scriptstyle \b \b$}}}
\def\gnnnn{\Gamma_4^{\hbox{$\scriptstyle \n \n$}\kern -8.2pt  
\raise -4pt \hbox{$\scriptstyle \n \n$}}}
\def\gnnnnnn{\Gamma_6^{\hbox{$\scriptstyle \n \n \n$}\kern -12.3pt
\raise -4pt \hbox{$\scriptstyle \n \n \n$}}}
\def\gbbbbbb{\Gamma_6^{\hbox{$\scriptstyle \b \b \b$}\kern -12.3pt
\raise -4pt \hbox{$\scriptstyle \b \b \b$}}}
\def\gbbbbc{\Gamma_{4\, c}^{\hbox{$\scriptstyle \b \b$}\kern -8.2pt
\raise -4pt \hbox{$\scriptstyle \b \b$}}}
\def\gnnnnc{\Gamma_{4\, c}^{\hbox{$\scriptstyle \n \n$}\kern -8.2pt
\raise -4pt \hbox{$\scriptstyle \n \n$}}}
\def\Rbud#1{{\cal R}_{#1}^{-\kern-1.5pt\blacktriangleright}}
\def\Rleaf#1{{\cal R}_{#1}^{-\kern-1.5pt\vartriangleright}}
\def\Rbudb#1{{\cal R}_{#1}^{\circ\kern-1.5pt-\kern-1.5pt\blacktriangleright}}
\def\Rleafb#1{{\cal R}_{#1}^{\circ\kern-1.5pt-\kern-1.5pt\vartriangleright}}
\def\Rbudn#1{{\cal R}_{#1}^{\bullet\kern-1.5pt-\kern-1.5pt\blacktriangleright}}
\def\Rleafn#1{{\cal R}_{#1}^{\bullet\kern-1.5pt-\kern-1.5pt\vartriangleright}}
\def\Wleaf#1{{\cal W}_{#1}^{-\kern-1.5pt\vartriangleright}}
\def\Cleaf{{\cal C}^{-\kern-1.5pt\vartriangleright}}
\def\Cbud{{\cal C}^{-\kern-1.5pt\blacktriangleright}}
\def\Crleaf{{\cal C}^{-\kern-1.5pt\circledR}}


\magnification=\magstep1
\baselineskip=12pt
\hsize=6.3truein
\vsize=8.7truein
 at 8truept
 at 8truept
 at 10truept

\font\bigrm=cmr12 at 14pt
\centerline{\bigrm Mass distribution exponents for growing trees}

\bigskip\bigskip

\centerline{F. David${}^{1}$, P. Di Francesco${}^1$, E. Guitter${}^1$ and T. Jonsson${}^2$}
  \medskip
  \centerline{\it ${}^1$Service de Physique Th\'eorique, CEA/DSM/SPhT}
  \centerline{\it Unit\'e de recherche associ\'ee au CNRS}
  \centerline{\it CEA/Saclay}
  \centerline{\it 91191 Gif sur Yvette Cedex,}
  \centerline{\it France}
  \smallskip
  \centerline{\it ${}^2$Science Institute,}
  \centerline{\it University of Iceland,}
  \centerline{\it Dunhaga 3, 107 Reykjavik,}
  \centerline{\it Iceland}
  \medskip
\centerline{\tt francois.david@cea.fr}
\centerline{\tt philippe.di-francesco@cea.fr}
\centerline{\tt emmanuel.guitter@cea.fr}
\centerline{\tt thjons@raunvis.hi.is}

  \bigskip
  \bigskip
  \bigskip


     \bigskip\bigskip

     \centerline{\bf Abstract}
     \smallskip
     {\narrower\noindent
We investigate the statistics of trees grown from some initial tree
by attaching links to preexisting vertices, with attachment probabilities 
depending only on the valence of these vertices. We consider the 
asymptotic mass distribution that measures the repartition of the mass 
of large trees between their different subtrees. This distribution is shown 
to be a broad distribution and we derive  explicit expressions for scaling 
exponents that characterize its behavior when one subtree is much 
smaller than the others. We show in particular the existence of various regimes 
with different values of these mass distribution exponents. Our
results are corroborated by a number of exact solutions for 
particular solvable cases, as well as by numerical simulations.
\par}

   \bigskip


\nref\book{J.~Ambj\o rn, B.~Durhuus and T.~Jonsson, {\it Quantum
geometry: a statistical field theory approach,} Cambridge University Press,
Cambridge (1997).}

\nref\AL{J. Ambj\o rn and R. Loll, {\it Non-perturbative Lorentzian
Quantum Gravity, Causality and Topology Change},
Nucl. Phys. {\bf B536 [FS]} (1998) 407, hep-th/9805108.}

\nref\LORGRA{P. Di Francesco and E. Guitter, {\it Critical and Multicritical Semi-Random $(1+d)$-
Dimensional Lattices and Hard Objects in $d$ Dimensions}, J.Phys. A35 (2002) 897-928.}

\nref\albert{R.~Albert and A.-L.~Barab\'{a}si, {\it Statistical mechanics
of complex networks}, Rev.\ Mod.\ Phys. {\bf 74} (2002) 47-97.}

\nref\specdim{B.~Durhuus, T.~Jonsson and J.~F.~Wheater, {\it The spectral
dimension of generic trees}, [math-ph/0607020].}

\nref\bergfinnur{B.~Durhuus, {\it Probabilistic aspects of infinite trees
and surfaces,} Act. Phys. Pol. {\bf 34} (2003) 4795-4811.}

\nref\pemantle{see R.~Pemantle, {\it A survey of random processes with
reinforcement} and references therein, [math.PR/0610076].}

\nref\janson{S.~Janson, {\it Functional limit theorems for multitype
branching processes and generalized P\'olya urns}, Stochastic Processes
and their Applications {\bf 110} (2004) 177-245.}

\nref\drmota{M.~Drmota, S.~Janson and R.~Neininger, {\it A functional
limit theorem for the profile of search trees}, [math.PR/0609385].}

\nref\rudas{A.~Rudas, B.~Toth and B.~Valko, {\it Random trees and general
branching processes}, [math.PR/0503728].}

%
%
%
%
\vfill
\eject
\newsec{Introduction}
Random trees arise in many branches of science, ranging from the social
sciences through biology, physics and computer science to pure mathematics.
In physics, random trees often occur in the context of statistical
mechanics or quantum field theory
where the weight or probability of a tree is given by a local
function, e.g.\ a Boltzmann factor with an energy which depends only on the
valence of individual vertices but not on global features of the tree. A
motivation for study of such trees comes for instance from results in 
the theory of random surfaces which behave like trees in some cases, see \book.
Another motivation concerns the study of two-dimensional quantum gravity 
by use of so-called causal tessellations \AL, directly expressible in 
terms of random walks or of trees \LORGRA.

Another large class of random trees arises in {\it growth processes}
 where a tree evolves in time
by adding new vertices which attach themselves by a link to one of the
vertices of a preexisting tree according to some stochastic
rules, with local attachment probabilities. A possible realization is
a tree-like molecule (branched polymer) growing in a solution with an
abundance of monomers.  
In many important real world cases one can observe the structure
of the growing tree but only
guess the rules which govern the growth (social networks, the internet,
citation networks etc.). For a review of this topic, see \albert.

Most tree models in statistical mechanics with a local energy function fall
into one universality class, called {\it generic random trees}, with
susceptibility exponent $\gamma =1/2$, intrinsic Hausdorff dimension $2$ 
and spectral dimension $4/3$ \specdim.  This universality class corresponds
also to the mean field theory of branched polymers. If we look at a generic 
infinite tree, then with probability $1$ there is a unique infinite 
non-backtracking
path \bergfinnur.  A rooted generic tree can therefore be viewed as an
infinite half-line with finite trees growing out of it and these finite
outgrowths have a well-understood distribution.  This means that if we sit
at a vertex of a generic random tree, almost all the vertices of the tree
are to be found in the direction of one of the links emanating from this
vertex.

There do not seem to be any local growth rules for a tree which produce
generic trees.
In this paper we study the structure of growing trees and
focus on how they differ from generic trees.  Taking binary planar trees with a
root of valence one as an example we can ask what proportion of the vertices
sits on the right hand side of the tree and what proportion on the left.  For
a generic tree almost all the vertices are on one side.  We will see that with
local growth rules, in the limit of infinite trees, we get a continuous
distribution for the proportion of
vertices sitting on each side.  We will refer to this distribution as the
{\it mass distribution}. This quantity is one of the simplest geometrical
characterizations of the geometry of the random trees and gives a first
glimpse at their fractal structure. It might also be of practical use,
for instance when applied to binary search trees, to improve algorithmic
efficiency by anticipating the structure of stored data.
The aim of this paper is an investigation of this mass distribution, 
in particular via the derivation of scaling exponents characterizing   
its behavior when one side of the tree is much smaller than the other.
Those exponents turn out to depend precisely on the underlying growth
dynamics and also on the initial condition of the growth process.

As long as we are interested only in global properties such as
the mass distribution, the growth process reduces to a model of
evolution of populations which belongs to the class of so-called 
generalized P\'olya urn problems \pemantle. These problems have been analyzed
using continuous time branching processes [\xref\janson-\xref\rudas] and we
believe that the mass distribution exponents can in principle 
be deduced using those methods.  However, we find it illuminating to 
study the model directly using discrete time and elementary methods.

We first
consider growing binary trees which are characterized by a single parameter
$x$ given by the ratio between the probabilities that a new vertex
attaches itself to an old vertex of valence 1 or of valence 2.  For a certain
value of $x$ the model can be mapped onto a simple model of reinforced
random walk which is exactly solvable. When starting from the 
tree consisting of a single link, one finds a uniform
mass distribution, i.e.\ the probability that a fraction $u$ of the total
mass sits on the left is independent of $u$. More generally, this distribution 
is given by a Beta law that depends
on the initial tree that we start the growth process from. 
For other, arbitrary values of the parameter $x$, we show how to 
calculate the mass 
distribution exponents exactly by scaling arguments. Again, these
exponents depend strongly on the initial condition.

In the second half of the paper these results are generalized to the case of
growing trees with vertices of a uniformly bounded valence.  Both for binary
and the more general multinary trees
numerical experiments have been carried out and are in excellent agreement
with the scaling results.

\newsec{Growth of binary trees}

\subsec{Definition of the model}
\fig{A sample binary tree (top, thick lines) grown on top of an underlying infinite binary
Cayley tree (thin lines) by attaching a new link either at a univalent vertex (bottom left)
or at a bivalent one (bottom right). The choice of the attachment vertex is made with 
probability weights $w_i$ depending only on its valence $i=1$ or $2$. When $i=1$, the new link
is chosen among the two possible descendants with equal probabilities.}{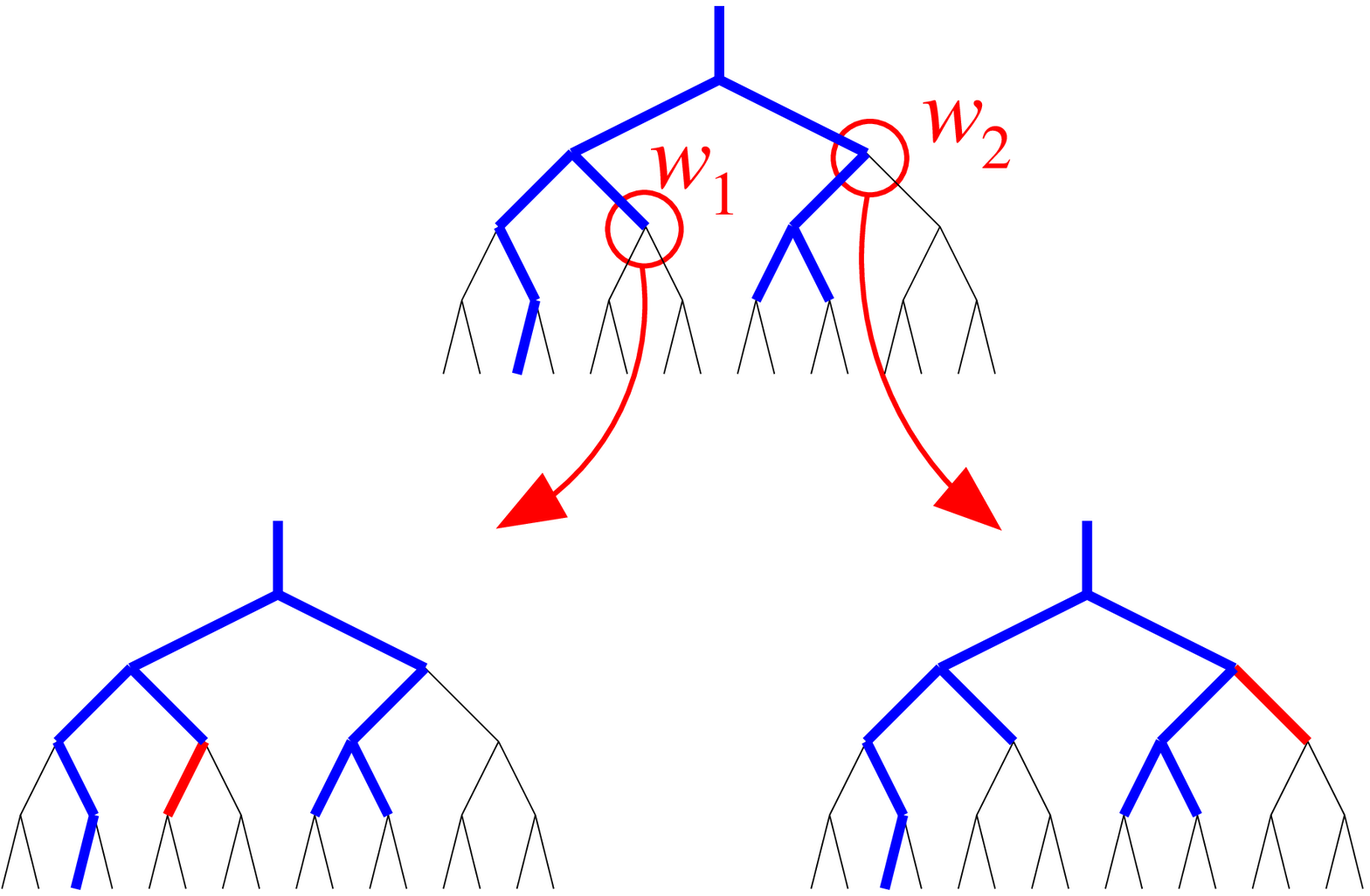}{10.cm}
\figlabel\binary
The model that we will study consists of growing binary trees from some finite  initial binary tree
${\cal T}^0$ by successive addition of links. 
As usual, the binary trees are rooted at some {\it root edge} and planar, which allows to distinguish 
left and right descendants at each vertex. 
One may alternatively think of the trees as drawn on top of some underlying 
infinite binary Cayley tree with a unique leaf connected to the {\it root edge} (see figure \binary). 
The growing of a new link takes place at uni- or bi-valent vertices of the existing 
tree only, with two choices (left or right descendant) in the first case and a unique choice (the
descendant not already occupied) in the second case. At each step, the choice of the vertex to which 
we attach the new link is made with a probability that depends on its valence $i$ ($i=1,2$)
and chosen to be proportional to $w_1$ for univalent vertices and $w_2$ for bivalent ones. 
More precisely, if we denote by $n_i$, $i=1,2,3$ the total number of $i$-valent vertices on the growing 
tree at a given step,  the probability to attach the new link to a given $i$-valent vertex reads 
$w_i/(n_1 w_1+n_2 w_2)$ for $i=1,2$.  Note that the process depends only on the quantity $w_1/w_2$.
Once the choice of attachment vertex is made, the precise choice of link to add is then unambiguous 
if $i=2$ and uniformly distributed on the left and right descendants if $i=1$. 

In the following, we will concentrate on {\it global properties} of the growing tree, such as its
total numbers $n_i$ of $i$-valent vertices. The above growing process induces a Markovian
evolution for $n_i(t)$ as functions of a {\it discrete time} $t$ equal to the number of added links,
starting from $n_i(0)\equiv n_i^0$, the numbers of $i$-valent vertices on ${\cal T}^0$.
We have the following evolution rules:
\eqn\rules{\eqalign{
{\rm with\ probability\ } {n_1(t) w_1\over n_1(t)w_1+n_2(t)w_2}\ , \ 
\quad\left\{\matrix{n_1(t+1)&=n_1(t)\hfill\cr n_2(t+1)&= n_2(t)+1\cr n_3(t+1)&=n_3(t)\hfill\cr}\right.\cr
{\rm with\ probability\ } {n_2(t) w_2\over n_1(t)w_1+n_2(t)w_2}\ , \
\quad\left\{\matrix{n_1(t+1)&=n_1(t)+1\cr n_2(t+1)&= n_2(t)-1\cr n_3(t+1)&=n_3(t)+1\cr}\right.\cr}}
These rules are a particular realization of so-called generalized P\'olya
urns with three types of ``balls" taken out of a single urn. The problem
can thus be studied by continuous time branching processes (see \janson)
but we prefer to present here some heuristic but more direct arguments.   
The rules \rules translate into a {\it master equation} for the 
probability $p(n_1,n_2,n_3;t)$ that there 
be $n_i$ $i$-valent vertices in the growing tree at time $t$:
\eqn\master{\eqalign{p(n_1,n_2,n_3;t+1)&=  {n_1 w_1\over n_1w_1+(n_2-1)w_2} \ p(n_1,n_2-1,n_3;t)
\cr &+
 {(n_2+1) w_2\over (n_1-1)w_1+(n_2+1)w_2} \ p(n_1-1,n_2+1,n_3-1;t)\ .}}
This, together with the initial conditions $p(n_1,n_2,n_3,0)= \prod\limits_{i=1}^3 \delta_{n_i,n_i^0}$,
determines $p(n_1,n_2,n_3;t)$ completely.

Defining the total {\it mass} $T$ of the evolving tree as its total number of edges (including the root edge), 
we have clearly the relation $T=t+T^0$, where $T^0$ is the mass of 
the initial tree ${\cal T}^0$. Moreover, we have
the standard relations for binary trees:
\eqn\mass{T=n_1+n_2+n_3=2n_1+n_2-1\ .}
We may therefore write  
\eqn\conserv{p(n_1,n_2,n_3;t)= \delta_{n_2,t+T^0+1-2n_1}\ \delta_{n_3,n_1-1}\  p(n_1;t)\ .}
Hence, the problem reduces to finding a function of two variables only.

The proportions $n_i(t)/T$ of the total mass corresponding to edges antecedent of $i$-valent vertices
provide a first set of interesting quantities. As we shall see in the next section, these proportions
tend asymptotically at large $t$ to fixed values depending on $w_1/w_2$ only and {\it not} 
on the initial condition. Moreover, these values may be computed exactly via a simple mean-field 
argument.

As mentioned in the introduction, 
another interesting quantity that characterizes the geometry of the
growing tree is given by the repartition of the total mass between {\it the 
left and the right of the root edge}. Defining the left and right masses $T_L$ and $T_R$ as 
the total number of edges in the left and right descending subtrees of the root edge 
(with $T_L+T_R=T-1$), we shall be interested in the asymptotic proportions $u_L=T_L/(T-1)$ 
and $u_R=T_R/(T-1)$ at large $t$ (with $u_R=1-u_L$).
The corresponding limiting law $P(u_L,u_R)$ is quite subtle and cannot be obtained 
by mean-field arguments. As we shall see below, 
the proportions $u_L$ and $u_R$ are not peaked to fixed values but are instead characterized by a broad
asymptotic probability distribution $P(u_L,u_R)$ on $[0,1]$, which moreover depends strongly on the initial 
conditions.
The main subject of this paper is an investigation of this limiting law and in particular the
derivation of exponents that characterize it.

\subsec{Mean-field results}

The mean-field approach consists in assuming that the proportions of mass under study
are peaked at large $t$ around fixed values and in neglecting fluctuations around those.
As we shall see, this gives a fully consistent result in the case of the proportions
$n_i/T$ above, while it gives no information on the left and right proportions
$u_L$ and $u_R$. 

Multiplying both sides of eq. \master\ by $n_1$, (resp. $n_2$, $n_3$) and summing over
all $n$'s at fixed $t$, we get
\eqn\aver{\eqalign{ 
\langle n_1\rangle_{t+1} &= \langle n_1\rangle_t +
\langle {n_2 w_2 \over n_1 w_1 +n_2 w_2} \rangle_t \cr
\langle n_2\rangle_{t+1} &= \langle n_2\rangle_t +
\langle {n_1 w_1 -n_2 w_2 \over n_1 w_1 +n_2 w_2} \rangle_t \cr
\langle n_3\rangle_{t+1} &= \langle n_3\rangle_t +
\langle {n_2 w_2 \over n_1 w_1 +n_2 w_2} \rangle_t \cr
}}
Here, the notation $\langle \cdots \rangle_t$ stands for the average over all possible growths
of the initial tree ${\cal T}^0$ by adding $t$ links. 
The mean-field hypothesis allows to substitute $n_i \to \alpha_i t$ within averages in \aver,
with constants $\alpha_i$ to be determined. This substitution immediately yields
\eqn\eqalpha{\alpha_1=\alpha_3={\alpha_2 w_2 \over \alpha_1 w_1+\alpha_2 w_2} \qquad
\alpha_2={\alpha_1 w_1-\alpha_2 w_2 \over \alpha_1 w_1+\alpha_2 w_2}}
which fixes
\eqn\valalpha{\alpha_1=\alpha_3={2\,x\over 3\,x+\sqrt{x(8+x)}}\qquad
\alpha_2={2\over 2+x+\sqrt{x(8+x)}}}
where
\eqn\defx{x\equiv{2 w_2\over w_1}\ .}
\fig{The limiting proportions $\alpha_i\sim n_i/t$ at large $t$ as functions of the ratio $x=2w_2/w_1$, 
with values \valalpha\ predicted by a mean field argument.}{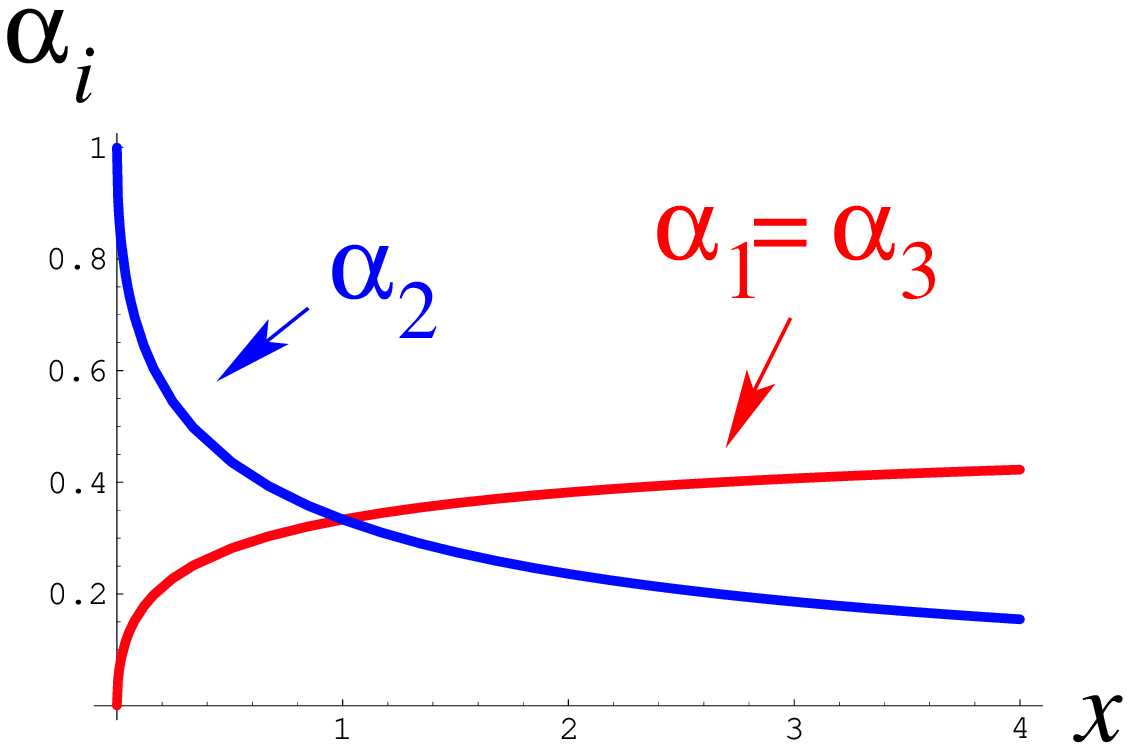}{6.cm}
\figlabel\proportions
\noindent These values are plotted in fig.\proportions.
Note that we have the relations $\alpha_1+\alpha_2+\alpha_3=2\alpha_1+\alpha_2=1$ in agreement
with eq. \mass.

To check this mean-field result, we may use the master equation \master\ to estimate
the asymptotics of $p(n_1,t)$ in \conserv. Performing the substitution \conserv\ into
eq. \master, we get
\eqn\masterbis{\eqalign{p(n_1;t+1)&=  {n_1 w_1\over n_1w_1+(t+T^0+1-2 n_1)w_2} \ p(n_1;t)
\cr &+
 {(t+T^0+3-2 n_1) w_2\over (n_1-1)w_1+(t+T^0+3-2n_1)w_2} \ p(n_1-1;t)}}
At large $t$, we assume that the solution takes the asymptotic form
\eqn\asympform{p(n_1;t)\sim {1\over \sqrt{t}}f\left( {n_1\over \sqrt{t}}-\alpha_1 \sqrt{t}\right)}
for some function $f(z)$ and some parameter $\alpha_1$ to be determined. 
Substituting this form into eq. \masterbis\ and expanding 
in $1/\sqrt{t}$ at large $t$, we get at leading order the consistency relation
\eqn\consistency{{w_1\over w_2}={(1-\alpha_1)(1-2\alpha_1)\over \alpha_1^2}}
from which we recover the value $\alpha_1$ of eq. \valalpha, while the values 
of $\alpha_2$ and $\alpha_3$ follow from $\alpha_2=1-2\alpha_1$ and $\alpha_3=\alpha_1$.
At sub-leading order, we get a differential equation for $f(z)$:
\eqn\diffeq{(3-4\alpha_1)(f+z\, f')+\alpha_1(1-\alpha_1)(1-2\alpha_1)f''=0}
from which we deduce that
\eqn\valf{f(z)\propto e^{-{3-4\alpha_1\over 2 \alpha_1(1-\alpha_1)(1-2\alpha_1)}z^2}}
As announced, the asymptotic distribution for $n_1/t$ is peaked
around the mean value $\alpha_1$. More precisely, it is Gaussian with 
a width of order $1/\sqrt{t}$, hence tends
to $\delta(n_1/t-\alpha_1)$ when $t\to \infty$.

It is interesting to apply the same mean-field approach to the case of the left-right repartition
of the mass in the growing tree. The above Markov process must now be refined so as to keep track
of the left (L) and right (R) numbers of uni- and bi-valent vertices $n_{i,A}$, $i=1,2$ and
$A=L,R$. The masses on the left and right of the root edge are then simply expressed 
as $T_L=2n_{1,L}+n_{2,L}-1$ and $T_R=2n_{1,R}+n_{2,R}-1$. Here and throughout the paper, we
use the convention that $n_{1,L}=0$, $n_{2,L}=1$ if the left subtree is empty, and similarly for the
right subtree. If we now assume the existence of limiting
proportions $n_{i,A}/t\to \eta_{i,A}$ at large $t$, we get the mean-field equations
\eqn\meaneq{\eqalign{
\eta_{1,L}&=w_2 \eta_{2,L}/\Sigma \cr
\eta_{2,L}&=(w_1 \eta_{1,L}-w_2 \eta_{2,L})/\Sigma \cr
\eta_{1,R}&=w_2 \eta_{2,R}/\Sigma \cr
\eta_{2,R}&=(w_1 \eta_{1,R}-w_2 \eta_{2,R})/\Sigma \cr}}
where $\Sigma=\Sigma_L+\Sigma_R$ and $\Sigma_A=w_1 \eta_{1,A}
+w_2 \eta_{2,A}$. We introduce the quantities $u_A\equiv
2 \eta_{1,A}+\eta_{2,A}$, which are nothing but the asymptotic
proportions of the total mass lying on the left and right of the root edge.
{}From eq. \meaneq, we deduce that $u_A=\Sigma_A/\Sigma$, henceforth
$u_L+u_R=1$ as it should. Introducing $\alpha_{i,A}=\eta_{i,A}/u_A$,
which are the proportions of uni-and bi-valent vertices {\it within} the left, resp. right
subtree, eqs. \meaneq\ decouple into two sets of equations
\eqn\meaneqbis{\eqalign{ & \left\{\matrix{
\alpha_{1,L}&=\alpha_{2,L}w_2/(\alpha_{1,L}w_1+\alpha_{2,L}w_2) \hfill\cr
\alpha_{2,L}&=(\alpha_{1,L}w_1-\alpha_{2,L}w_2)/(\alpha_{1,L}w_1+\alpha_{2,L}w_2)\hfill \cr} \right.\cr
& \left\{\matrix{
\alpha_{1,R}&=\alpha_{2,R}w_2/(\alpha_{1,R}w_1+\alpha_{2,R}w_2)\hfill \cr
\alpha_{2,R}&=(\alpha_{1,R}w_1-\alpha_{2,R}w_2)/(\alpha_{1,R}w_1+\alpha_{2,R}w_2)\hfill \cr} \right.\cr }}
which are identical to the mean field equation \eqalpha. We deduce that
$\alpha_i^{L}=\alpha_i^{R}=\alpha_i$ of eq. \valalpha\ while $u_L$ and $u_R$
remain {\it undetermined}. In other words, the relative proportions of uni- and bi-valent
vertices within each subtree tend asymptotically to the same mean field values as in the whole
tree, but the distribution $P(u_L,u_R)$ is not peaked to particular values. We expect instead
a broad distribution in the whole interval $[0,1]$. This property will be illustrated
in the next section in a particular case.

\subsec{A solvable case}

A particularly interesting case corresponds to growing the binary trees by adding links
chosen {\it uniformly at random} among all possible available positions. This amounts to taking
$w_1=2 w_2$ as there are twice as many available descendants at univalent vertices as
there are at bivalent ones. This corresponds to $x=1$ in \defx. For simplicity, we take
$w_1=2$ and $w_2=1$ so that $w_1 n_1+w_2 n_2 = 2 n_1+n_2 $ directly counts the number of available positions
for the addition of a link. This number is nothing but the mass $T$ of the tree plus $1$.
Therefore, the evolution process no longer depends on $n_1$ and $n_2$ individually, but instead
depends on the total mass $T$ only. 

Turning to left and right proportions, we need only keep track
of the masses $T_L$ and $T_R$ of the right and left subtree (with $T_L+T_R=T-1$). The induced 
evolution process consists at each step in increasing the mass by one on the left with 
probability $(T_L+1)/(T+1)$ and on the right with probability  $(T_R+1)/(T+1)$.
This process is known as the reinforced random walk (RRW) in which a walker goes to the
left (resp. to the right) with a probability proportional to the number of times he has already
stepped in this direction. For a review on reinforced processes, see
for instance \pemantle.
We may easily write a master equation for the probability
$p(T_L,T_R;t)$ to have masses $T_L$ and $T_L$ at time $t$
(with $T=t+T^0$):

\eqn\mastersolv{p(T_L,T_R;t+1)= {T_L\over t+T^0+1}p(T_L-1,T_R;t)+{T_R\over t+T^0+1}p(T_L,T_R-1;t)} 
where $T^0=T^0_L+T^0_R+1$ and $T^0_L$ and $T^0_R$ are the initial left and right masses.
Together with the initial condition $p(T_L,T_R,0)=\delta_{T_L,T_L^0}\delta_{T_R,T_R^0}$,
this fixes the solution to be
\eqn\solvable{p(T_L,T_R;t)={{T_L\choose T_L^0}{T_R\choose T_R^0}\over
{t+T^0\choose T^0}}\ \delta_{T_L+T_R+1,t+T^0}}
In particular, when we start from an initial tree ${\cal T}^0$ consisting of 
the root edge alone, we
have $T^0=1$, $T_L^0=T_R^0=0$ and therefore $p(T_L,T_R;t)= \delta_{T_L+T_R,t}/(t+1)$.
Hence, all the values $T_L=0,1,\ldots,t$ are {\it equiprobable}. This is a well-known
feature of the RRW. 

At large $t$ and for arbitrary initial conditions, we may extract from eq. \solvable\ the
asymptotic law for the probability distribution of 
the left and right mass proportions $u_L$ and $u_R$ by writing 
$P(u_L,u_R)=\lim\limits_{t\to  \infty}t^2\ p(u_L(t+T^0-1),u_R(t+T^0-1);t)$.
We get explicitly
\eqn\asympsol{P(u_L,u_R)={(T_L^0+T_R^0+1)!\over T_L^0!\, T_R^0!}\  u_L^{T_L^0}\  u_R^{T_R^0}\  
\delta(u_L+u_R-1)}
We thus get a simple Beta law with exponents $T_L^0$ and $T_R^0$, the left and right masses of
the initial tree.

At this stage, a few remarks are in order. First, as announced, the distribution $P(u_L,u_R)$ is
supported over the whole interval $[0,1]$. Next, it explicitly depends on the initial condition.
However, we note that this dependence is only through the initial masses and does not 
involve the precise shape of the initial left or right trees. In view of this discussion,
we define the {\it left and right exponents} $\beta_L$ and $\beta_R$ through
\eqn\defexp{\eqalign{
P(u_L,u_R)&\sim u_L^{\beta_L}\ \ \  \hbox{when $u_L\to 0$}\cr
P(u_L,u_R)&\sim u_R^{\beta_R}\ \ \ \hbox{when $u_R\to 0$}\cr}}
Here we have $\beta_L=T_L^0$ and $\beta_R=T_R^0$, and, as the distribution is a Beta law, 
the left and right exponents characterize it completely. Note that the left exponent $\beta_L$
depends only on the left initial condition and not on the right one and vice-versa.
As we will see later, this property will remain true for arbitrary values of $w_1$ and $w_2$.

\newsec{Mass distribution exponents}

In this section, we derive the left and right exponents $\beta_L$ and $\beta_R$ defined
in \defexp\ for the general case of arbitrary $w_1$ and $w_2$ and for arbitrary initial
conditions. 

\subsec{Results}

Before we proceed to the actual derivation of the exponents, let us display and discuss the 
corresponding formula that we obtain.
We have the simple expression
\eqn\genexp{\eqalign{
\beta_L&= -1+{1 \over \alpha_1 w_1 +\alpha_2 w_2}
\ \min\limits_{{\cal T}\supset {\cal T}_L^0} \{w_1 n_1({\cal T})+w_2 n_2({\cal T}) \} \cr
\beta_R&= -1+{1 \over \alpha_1 w_1 +\alpha_2 w_2}
\ \min\limits_{{\cal T}\supset {\cal T}_R^0} \{w_1 n_1({\cal T})+w_2 n_2({\cal T}) \} \cr
}}
where $\alpha_1$ and $\alpha_2$ are the limiting proportions given by eq. \valalpha\ 
and where the minimum is taken over all binary trees ${\cal T}$ containing the
left (resp. right) initial subtree ${\cal T}_L^0$ (resp. ${\cal T}_R^0$), 
with $n_i({\cal T})$ being the number of $i$-valent
vertices in ${\cal T}$. In the above formula, we use again the convention that whenever
${\cal T}$ is empty , we have $n_1({\cal T})=0$ and $n_2({\cal T})=1$. 
The minimum may be explicitly evaluated and depends on the relative values of $w_1$ and $w_2$. 
Indeed, any binary tree ${\cal T}$ containing, say ${\cal T}^0_L$, may be obtained 
from ${\cal T}^0_L$ by the successive addition of a number of links. At each step, the
quantity $n_1 w_1+n_2 w_2$ either increases by $w_2>0$ if the link is added at a univalent vertex,
or is shifted by $w_1-w_2$ if it is added at a bivalent vertex. If $w_1>w_2$, the quantity
$n_1 w_1+n_2 w_2$ may therefore only increase strictly and the minimum is attained 
for the initial state ${\cal T}={\cal T}_L^0$, with value $w_1 n_{1,L}^0+w_2 n_{2,L}^0$,
where $n_{i,L}^0\equiv n_i({\cal T}_L^0)$. When $w_1<w_2$, the minimum
is obtained by saturating each  bivalent vertex of ${\cal T}^0_L$ into a trivalent
vertex and a uni-valent one. The resulting tree ${\cal T}$ has no more bivalent vertices
and a number $n_{1,L}^0+n_{2,L}^0$ of univalent ones. The associated minimum now
reads $w_1(n_{1,L}^0+n_{2,L}^0)$. We finally obtain the following explicit expression:
\eqn\lutfin{\beta_L=-1+{x+\sqrt{x(8+x)}\over 8\,x} \ \big(4 n_{1,L}^0+(x+2-|x-2|) n_{2,L}^0\big)}
and a similar expression for the right exponent. As mentioned above, we use the convention that
$n_{1,L}^0=0$ and $n_{2,L}^0=1$ if ${\cal T}_L^0$ is empty. Note that the left exponent only depends 
on the left initial subtree via its numbers $n_{i,L}^0$ of uni- and bi-valent vertices. 
\fig{Left mass distribution exponent $\beta_L$ as a function of $x=2w_2/w_1$ for an initial empty left
subtree, with value given by \lutfin\ with $n_{1,L}^0=0$ and $n_{2,L}^0=1$. At $x=1$, we see the
value $\beta_L=0$ of section 2.3. A change of regime takes place at $x=2$, i.e.
$w_1=w_2$.}{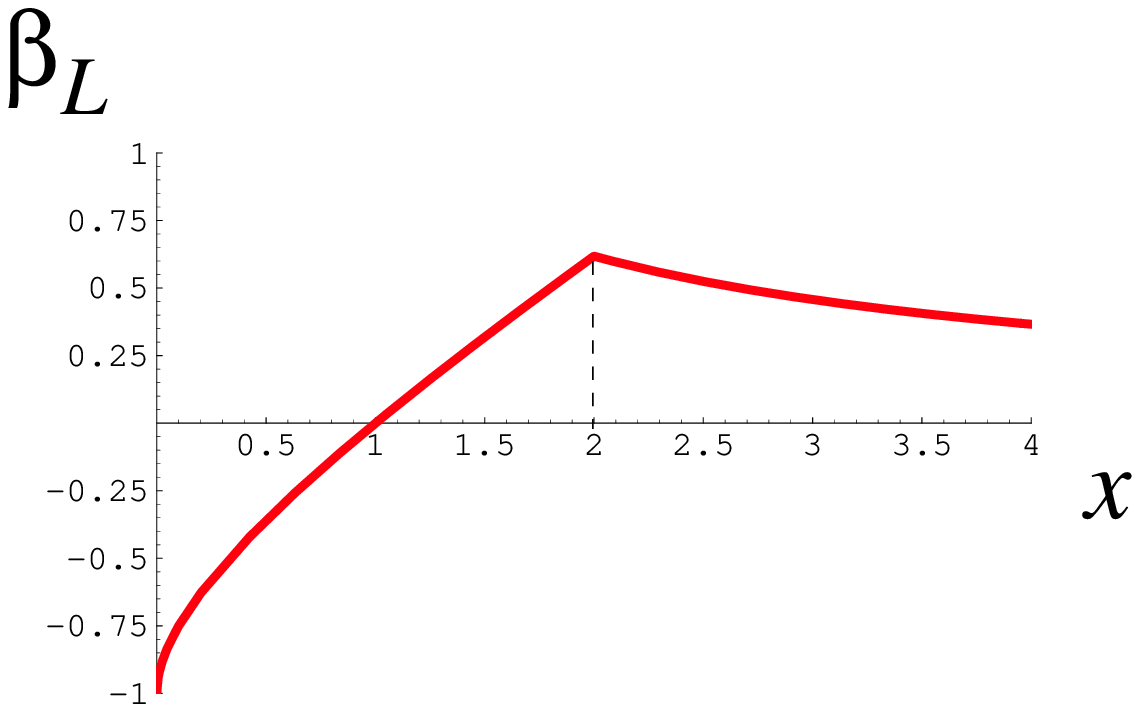}{7.cm}
\figlabel\exponent
The exponent $\beta_L$ is displayed in fig.\exponent\ for the simple case where we start from 
an empty left subtree. 

A few remarks are in order. First we note that, as a first check of our formula, we recover when $x=1$
the result $\beta_L=2 n_{1,L}^0+n_{2,L}^0-1=T_L^0$ of section 2.3. Two other particular values 
of $x$ may be easily checked, namely $x\to0$ and $x\to \infty$, as the corresponding
limiting growth processes may be easily analyzed.

When $x\to 0$, we find that $\beta_L$ tends to infinity unless $n_{1,L}^0=0$, which corresponds 
to an empty initial left subtree in which case $\beta_L=-1$. These results may be understood as
follows. For $x\to 0$, i.e. $w_1 >> w_2$, the trees that are built are extensions of
the initial tree by polymer-like chains (without new branching) attached to the
initial leaves. If at least one of the left or right initial subtrees is non-empty,
the number of attachment points for the chains is $n_{1,L}^0$ on the left
and $n_{1,R}^0$ on the right, with $n_{1,L}^0+n_{1,R}^0>0$, hence at large $t$, we have 
\eqn\llimitP{P(u_L,u_R)\to 
\delta(u_L-{n_{1,L}^0\over n_{1,L}^0+n_{1,R}^0})\ 
\delta(u_R-{n_{1,R}^0\over n_{1,L}^0+n_{1,R}^0})}
If $n_{1,L}^0=0$, we have a delta peak at $u_L=0$, which is consistent with the limiting
value $\beta_L=-1$. If $n_{1,L}^0>0$, we have delta peak at a positive value of $u_L$, 
consistent with a limiting value $\beta_L=\infty$. Finally, when both initial 
subtrees are empty ($n_{1,L}^0=n_{1,R}^0=0$), the first step consist in attaching a link
to the root edge with probability $1/2$ on the left or on the right and one then grows
a single chain on this side. We therefore have $P(u_L,u_R)=\big(\delta(u_L)\delta(u_R-1)
+\delta(u_L-1)\delta(u_R)\big)/2$, again consistent with $\beta_L=-1$.
\fig{A typical growth process at $x=\infty$ (from left to right). The process starts by
first saturation all bivalent vertices of the initial tree ${\cal T}^0$ (left) into trivalent ones,
resulting in a tree ${\cal S}^0$ (middle) with no more bivalent vertices. The growth then
proceeds by a two-step elementary process which consist in adding the two descendant links
to a leaf (circled) chosen uniformly among all leaves, and so on.}{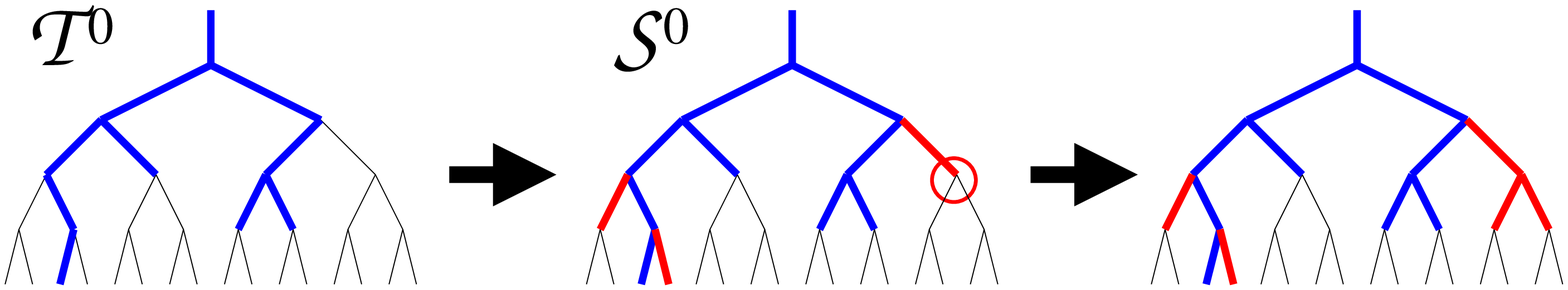}{13.cm}
\figlabel\satur
\fig{The bijection between ``saturated" binary trees (top left) without
bivalent vertices and ordinary binary trees (top right), obtained by erasing the root edge 
and squeezing all pairs of descendants into single links. The $x=\infty$ growth process
consisting in adding both descendants of a leaf chosen uniformly at random (bottom left) is bijectively
mapped onto the $x=1$ growth process where single links are added uniformly at random 
(bottom right).}{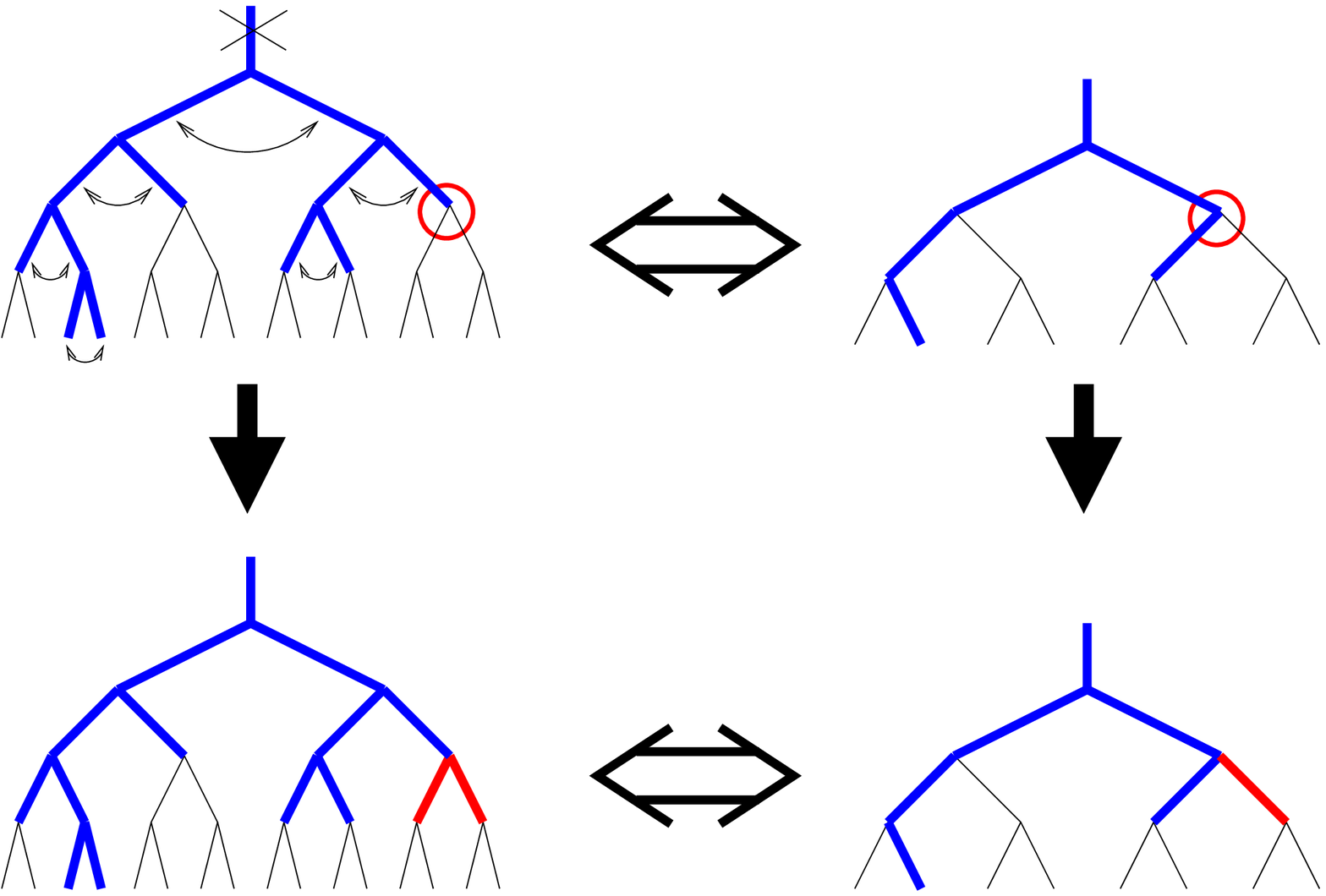}{10.cm}
\figlabel\bijec
Another interesting case is the limit $x\to \infty$, i.e. $w_2>>w_1$. 
In this case, the growth process consists in first saturating all initial bivalent vertices
into trivalent ones, resulting in a tree ${\cal S}^0$ without bivalent vertex and
with left and right masses $S_A^0=2(n_{1,A}^0+n_{2,A}^0)-1$, $A=R,L$ (see fig.\satur).  
Again we use the convention that $n_{1,A}=0$, $n_{2,A}=1$ if
the tree on side $A$ is empty. Note that if we start from a configuration ${\cal T}^0$ reduced
to the root edge, the first two steps of the growing process consist in adding to the root edge 
the two links that descend from it. We define for consistency ${\cal S}^0$ as this resulting tree,
with $S_L^0=S_R^0=1$. The growth then proceeds by a succession of two-step elementary processes 
that add at some existing leaf a {\it pair of links} on its left and right descendants.
The choice of the leaf at which we attach the new pair is moreover uniform among all
leaves. Remarkably, this two-step process with initial condition ${\cal S}^0$ is completely 
equivalent to the process of section 2.3 with $x=1$, where we were choosing {\it single} descendants 
uniformly.
This is due to the existence of a bijection between ``saturated" binary trees without bivalent
vertices and ordinary binary trees illustrated in fig.\bijec, obtained by erasing the root edge 
and squeezing all pairs of descendants into single links. The left mass of the 
image of ${\cal S}^0$ under the above bijection reads $(S_L^0-1)/2=n_{1,L}^0+n_{2,L}^0-1$. 
{}From the results of section 2.3, we thus get for the left exponent the value 
$\beta_L = n_{1,L}^0+n_{2,L}^0-1$, in agreement with \lutfin\ at $x\to \infty$.

Beyond the three solvable cases $x=0,1$ and $\infty$ above, we can deduce a 
number of qualitative results from our general formula \lutfin. 
First, the distribution is not in general a Beta law. Indeed, this
would require that the exponents for different initial conditions
differ by integers, which is not the case for generic $x$. Second, 
if we start from the tree consisting of a single edge, the distribution 
can be maximal at $u_L=u_R=1/2$ for $x>1$, or conversely, it can be 
maximal at $u_L=0$ or $u_R=0$ for $x<1$. Third, we see a change of 
behavior of the model at $x=2$ with a change of determination for the 
exponents which suggests the existence of a competition between 
different effective growth mechanisms. Finally, the fact that
the exponent $\beta_L$ increases with the size of the initial 
tree is a general feature of reinforced processes. Indeed, starting
from a large tree amounts to impose a preferential value of
the mass ratio, leading to a narrower distribution.

\subsec{Large time behavior}

In order to estimate the, say left mass distribution exponent $\beta_L$, we are
led to consider growth processes in which the left subtree {\it remains
finite} at large time $t$. In other words, we wish to estimate the large $t$ asymptotics 
of the probability $p({\cal T}^0,{\cal T}_L^*;t)$ to grow a tree in $t$ steps from some initial tree 
${\cal T}^0$ (with left subtree ${\cal T}_L^0$) to a final state made of a {\it fixed final 
left subtree} ${\cal T}_L^*$ of finite mass $T_L^*$ and an arbitrary right subtree, 
necessarily with a large mass of order $t$. This probability clearly decays
at large $t$ and we expect a power law behavior of the form
\eqn\powerp{p({\cal T}^0,{\cal T}_L^*;t)\sim t^{-\gamma({\cal T}^0,{\cal T}_L^*)}}
The computation of the exponent $\gamma({\cal T}^0,{\cal T}_L^*)$ will be a prerequisite
for that of the mass distribution exponent $\beta_L$. 

The probability $p({\cal T}^0,{\cal T}_L^*;t)$ is the sum of the probabilities of
all possible tree-growths of total size $t$ leading from ${\cal T}^0$ to a tree with left
subtree ${\cal T}_L^*$. It will prove convenient to divide any such process
into a first initial sub-process evolving up to some fixed time $t_0$
and a later process from $t_0$ to $t$. The value of $t_0$ is kept finite but 
is chosen to be large enough so that the mean-field approximation can be applied to
the (large enough) right subtree at each step of its evolution between $t_0$ and $t$.
We may classify all processes according to the intermediate configuration ${{\cal T}^0}'$
attained at time $t_0$. Note that ${{\cal T}^0}'$ necessarily has a left subtree 
${{\cal T}_L^0}'$ intermediate between ${\cal T}_L^0$ and ${\cal T}_L^*$. 
This allows us to rewrite 
\eqn\intermed{p({\cal T}^0,{\cal T}_L^*;t)=\sum_{{{\cal T}^0}'\atop
{\cal T}_L^0\subset {{\cal T}_L^0}' \subset {\cal T}_L^*}\pi({\cal T}^0,{{\cal T}^0}';t_0)
p({{\cal T}^0}', {\cal T}_L^*;t-t_0)}
with some probabilities $\pi({\cal T}^0,{{\cal T}^0}';t_0)$ that correspond to the
initial sub-process. These probabilities are a finite set of fixed positive numbers 
{\it independent of} $t$.

In turn, the second part of the process may be sorted according to the {\it induced}
left-subtree growth process. The latter corresponds to the successive addition of links to 
the left subtree ${{\cal T}_L^0}'$ until we reach ${\cal T}_L^*$. We will denote
by $G_L$ such a left-subtree growth process, characterized by the sequence of trees leading
from ${{\cal T}_L^0}'$ to ${\cal T}_L^*$. Beside the precise
sequence of links added, we also need to specify the time $t_j$ at which the $j$-th
link is added, with $j$ running from $1$ to $J=T_L^*-{T_L^0}'$ (where ${T_L^0}'$ is
the mass of ${{\cal T}_L^0}'$) and with $t_0<t_1<t_2<\cdots<t_J\leq t$. 
Between any two consecutive such times, the growth process only affects the right 
subtree. This allows us to rewrite 
\eqn\leftgrowth{p({{\cal T}^0}',{\cal T}_L^*;t-t_0)=\sum_{G_L;t_1,t_2,\ldots,t_J} p(G_L;t_1,
t_2,\ldots,t_J)}
where the sum extends over all left-subtree growth processes $G_L$ with fixed initial
and final conditions as above, supplemented by the sequence of times $t_j$. 
The probability $p(G_L;t_1,\ldots,t_J)$ of any given such process 
depends only on the sequence of the numbers $n_i^j\equiv n_{i,L}(t_j)$ 
of uni- and bi-valent vertices in the left subtree at time $t_j$ (i.e., for $j\geq 1$, 
just after the addition of the $j$-th link). As before, we adopt the convention
that $n_1^0=0$, $n_2^0=1$ when ${{\cal T}_L^0}'$ is empty. The probability 
$p(G_L;t_1,\ldots,t_J)$ can be estimated as follows:
{}from time $t_j$ to time $t_{j+1}-1$ (with the convention that $t_{J+1}=t$), 
the process only takes place on the right, hence has a probability
\eqn\probint{\prod_{\tau=t_j}^{t_{j+1}-1}{n_{1,R}(\tau)w_1+n_{2,R}(\tau)w_2 \over
(n_1^j+n_{1,R}(\tau))w_1+(n_2^j+n_{2,R}(\tau))w_2}\sim
\prod_{\tau=t_j}^{t_{j+1}-1} \left(1+ {n_1^j w_1+n_2^j w_2 \over 
\alpha_1 w_1+\alpha_2 w_2}\ {1\over \tau}\right)^{-1}}
where we have used the mean-field estimate $n_{i,R}(\tau)\sim \alpha_i \tau$ with $\alpha_i$
as in \valalpha. At time $\tau=t_{j+1}$ ($0\leq j\leq J-1$), the process takes places on the left,
hence has a probability $\sim (n_1^j w_1+n_2^j w_2)/((\alpha_1 w_1+\alpha_2 w_2)t_{j+1})$.
This results in 
\eqn\estpgl{p(G_L;t_1,\ldots,t_J)\sim 
\prod_{\tau=t_0}^{t_{1}-1} \left(1+{q_0\over \tau}\right)^{-1}\times{q_0\over t_1}\times
\prod_{\tau=t_1}^{t_{2}-1} \left(1+{q_1\over \tau}\right)^{-1}\times{q_1\over t_2}\times\cdots\times
{q_{J-1}\over t_J}\times
\prod_{\tau=t_J}^{t-1} \left(1+{q_J\over \tau}\right)^{-1}}
where we have used the notation
\eqn\defqj{q_j\equiv {n_1^j w_1 +n_2^j w_2 \over \alpha_1 w_1+\alpha_2 w_2}}
We may now fix the left-subtree growth process $G_L$, i.e. keep the {\it same} sequence of link 
additions, but sum over the times $t_j$, resulting in a probability $p(G_L)$ such that
\eqn\leftgrowthbis{p({{\cal T}^0}',{\cal T}_L^*;t-t_0)=\sum_{G_L} p(G_L)}
The corresponding 
nested sums over intermediate times are further approximated by nested integrals, with the
result 
\eqn\estim{p(G_L)\sim \int_{t_0<t_1<\ldots<t_J<t} \left(\prod_{j=1}^J {dt_j\over t_j}\, q_{j-1}\left({t_{j-1}\over t_j}
\right)^{q_{j-1}} \right)\ \left({t_{J}\over t}\right)^{q_J}
=\sum_{j=0}^J {q_j\over q_J} \left({t_0\over t}\right)^{q_j}
\prod_{i=0\atop i\neq j}^J {q_i\over q_i-q_j}}
Again the precision of this latest approximation is governed by the choice of $t_0$,
and can be arbitrarily accurate by picking a large enough $t_0$. 
The large $t$ behavior of $p(G_L)$ therefore reads
\eqn\estibis{p(G_L)\sim t^{-\gamma}\quad {\rm where}\quad \gamma
={-\min\limits_{j\in \{0,1,\ldots,J\}} q_j}\ .}
Introducing the notation
\eqn\defq{q({\cal T})\equiv {n_1({\cal T})w_1+n_2({\cal T})w_2 \over \alpha_1 w_1+\alpha_2 w_2}}
for any tree ${\cal T}$ with $n_i({\cal T})$ $i$-valent vertices, the minimum in \estibis\
is equivalently obtained as the minimum of $q({\cal T})$ when ${\cal T}$ runs over all the intermediate
trees of the growth process $G_L$, namely 
\eqn\estiter{p(G_L)\sim t^{-\min\limits_{{\cal T}\in G_L} q({\cal T})}\ .}
Returning to $p({{\cal T}^0}',{\cal T}_L^*;t-t_0)$, we read from the summation in eq. \leftgrowthbis\
that 
\eqn\estifour{p({{\cal T}^0}',{\cal T}_L^*,t-t_0) \sim t^{-\min\limits_{{{\cal T}_L^0}'\subset {\cal T}
\subset {\cal T}_L^*} q({\cal T})}}
with the minimum now taken over all trees intermediate
between ${{\cal T}_L^0}'$ and ${\cal T}_L^*$. Finally, from eq. \intermed, we have to sum over
all possible trees ${{\cal T}^0}'$ at time $t_0$, extending the range of trees in the minimum to all trees
intermediate between ${\cal T}_L^0$ and ${\cal T}_L^*$. This results in a behavior 
$p({\cal T}^0,{\cal T}_L^*;t)\sim t^{-\gamma({\cal T}^0,{\cal T}_L^*)}$ with 
\eqn\finestim{ \gamma({\cal T}^0,{\cal T}_L^*)=\min\limits_{{\cal T}_L^0\subset {\cal T}\subset
{\cal T}_L^*} q({\cal T})}
In practice, the exponent $\gamma({\cal T}^0,{\cal T}_L^*)$ depends only weakly on ${\cal T}_L^*$.
It proves useful to define a {\it generic} exponent 
\eqn\defgamgener{\gamma({\cal T}^0)\equiv \min\limits_{{\cal T}_L^0\subset {\cal T}} q({\cal T})}
with no upper bound on ${\cal T}$. In view of the discussion of section 3.1, the minimum is
reached for some finite tree ${\cal T}_{\rm min}$ (with ${\cal T}_{\rm min}={\cal T}_L^0$ if $w_1>w_2$ 
while, when $w_2>w_1$, ${\cal T}_{\rm min}={\cal S}^0_L$, the tree obtained by saturating 
each bivalent vertex of ${\cal T}_L^0$ into a trivalent vertex). We immediately deduce 
that $\gamma({\cal T}^0,{\cal T}_L^*)$ takes the generic value $\gamma({\cal T}^0)$ as soon 
as ${\cal T}_L^*$ contains ${\cal T}_{\rm min}$.  Moreover, if this is not the case, we necessarily 
have $\gamma({\cal T}^0,{\cal T}_L^*)\geq \gamma({\cal T}^0)$.

\subsec{Scaling argument}

The left exponent $\beta_L$ can be obtained from the exponent $\gamma({\cal T}^0)$
via a simple scaling argument. Indeed the distribution $P(u_L,u_R)$ for the left/right distribution
of the mass is obtained as
\eqn\limitlaw {P(u_L,u_R)=\lim\limits_{t\to  \infty}t^2\ p(u_L(t+T^0-1),u_R(t+T^0-1);t)}
where $p(T_L,T_R;t)$ is the probability to have reached after $t$ steps left and right masses $T_L$
and $T_R$ respectively. This probability implicitly depends on the initial condition 
${\cal T}^0$ (of total mass $T^0$). We have of course $T_L+T_R=t+T^0-1$ hence we may write
\eqn\redup{p(T_L,T_R;t)=p(T_L;t)\ \delta_{T_L+T_R,t+T^0-1}}
as well as 
\eqn\reduP{P(u_L,u_R)=P(u_L)\ \delta(u_R+u_R-1)}
where the two reduced functions are now related through
\eqn\redulimitlaw {P(u_L)=\lim\limits_{t\to  \infty}t\ p(u_L(t+T^0-1);t)}
The exponent $\beta_L$ characterizes the behavior of $P(u_L)$ at small $u_L$ via
\eqn\betaLP{P(u_L)\sim u_L^{\beta_L}\hbox{\ \ when $u_L\to 0$}}
At large $t$ and for large but finite $T_L$, we may according to \redulimitlaw\ and \betaLP\ 
estimate $p(T_L;t)$ via
\eqn\estiP{p(T_L;t)\sim {1\over t} P\left({T_L\over t}\right) \sim {1\over t} \left({T_L\over t}
\right)^{\beta_L}}
hence $p(T_L;t)$ has a power law decay $t^{-(\beta_L+1)}$. The above estimate holds 
in the range $1<<T_L<<t$. The probability $p(T_L;t)$ is the sum of the probabilities
$p({\cal T}^0,{\cal T}_L^*;t)$ over all final left subtree ${\cal T}_L^*$ of mass $T_L$.
{}From the results of previous section, each $p({\cal T}^0,{\cal T}_L^*;t)$ decays as
$t^{-\gamma(({\cal T}^0,{\cal T}_L^*)}$ and, as $T_L$ is large enough, some of
the ${\cal T}_L^*$'s will reach the (dominant) generic value $\gamma({\cal T}^0)$.
We deduce that
\eqn\finded{\beta_L+1=\gamma({\cal T}^0)}
with $\gamma({\cal T}^0)$ given by \defgamgener. Hence, the announced result \genexp.

\subsec{Numerical checks}

To corroborate the above results, we have made a number of numerical checks.
Those are based on random generation of large time Markov processes according to
\rules or its left/right refinement. All simulations are made of growth processes
with $t=4000$ steps. The number $N$ of runs ranges from $2\times 10^5$ to $10^8$ according
to the desired precision and various values of $x=2w_2/w_1$ are explored.
\fig{The measured distribution (see text) for $n_1/t$ (red) and $n_2/t$ (blue) and the
corresponding mean-field predictions from \valalpha.}{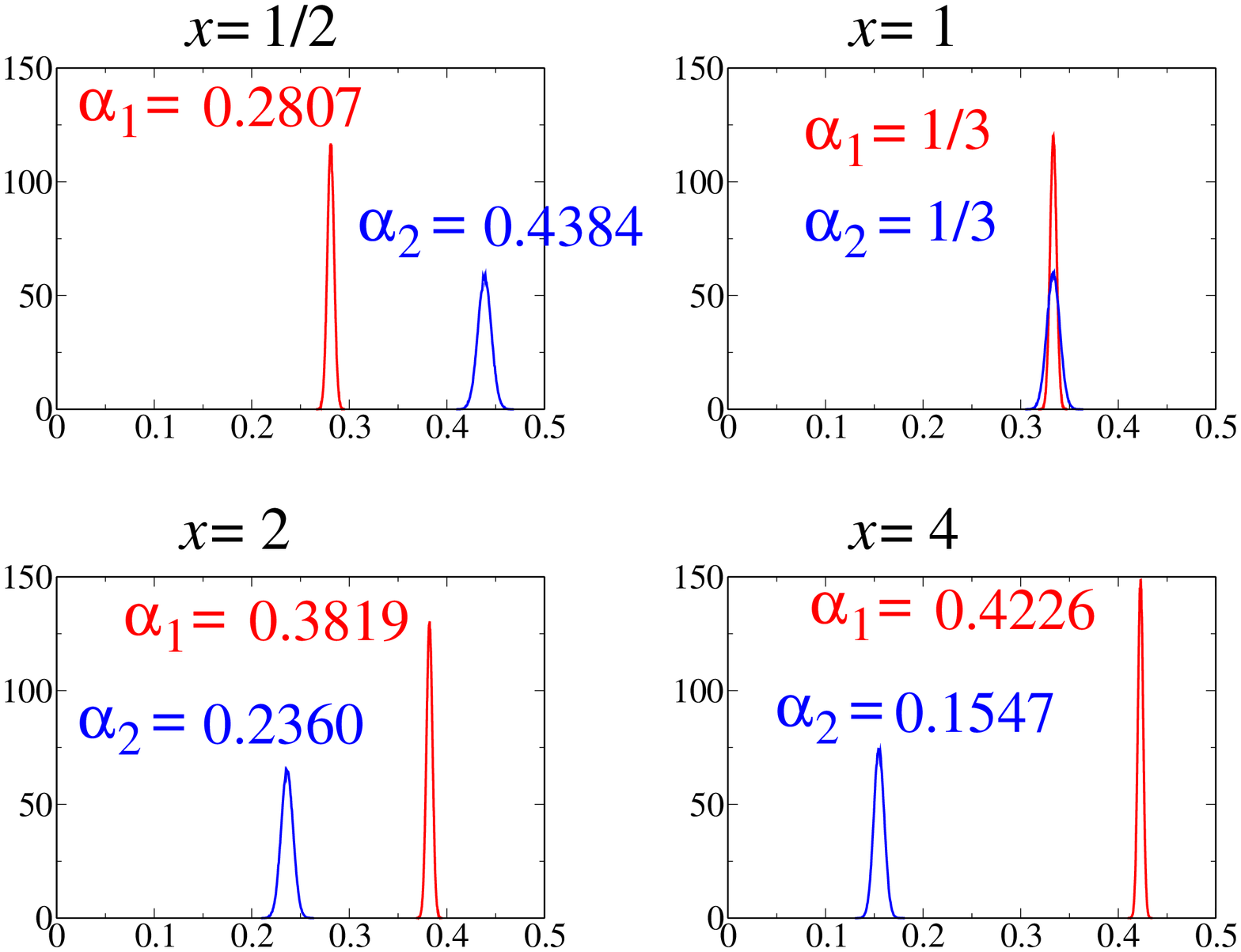}{12.cm}
\figlabel\alphacheck
The first test concerns the distribution of the ratios $n_i/t$ ($i=1,2$), to be
compared with the results of section 2.2. 
Figure \alphacheck\ displays the measured distribution obtained from $N=2\times 10^5$ runs
with an initial tree reduced to the root edge, and for $x=1/2,1,2$ and $4$. As expected,
we observe distributions peaked around precise values $\alpha_1$ and $\alpha_2$ corroborating
the mean field predictions \valalpha. The distributions and their widths agree perfectly
with the Gaussian shape \valf\ and its $\alpha_2$ counterpart.
\fig{Measured distribution $P(u_L)$ for the initial tree ${\cal T}^0$ drawn on the\nobreak\ left.
For $x=1$, we have also indicated the exact distribution. 
}{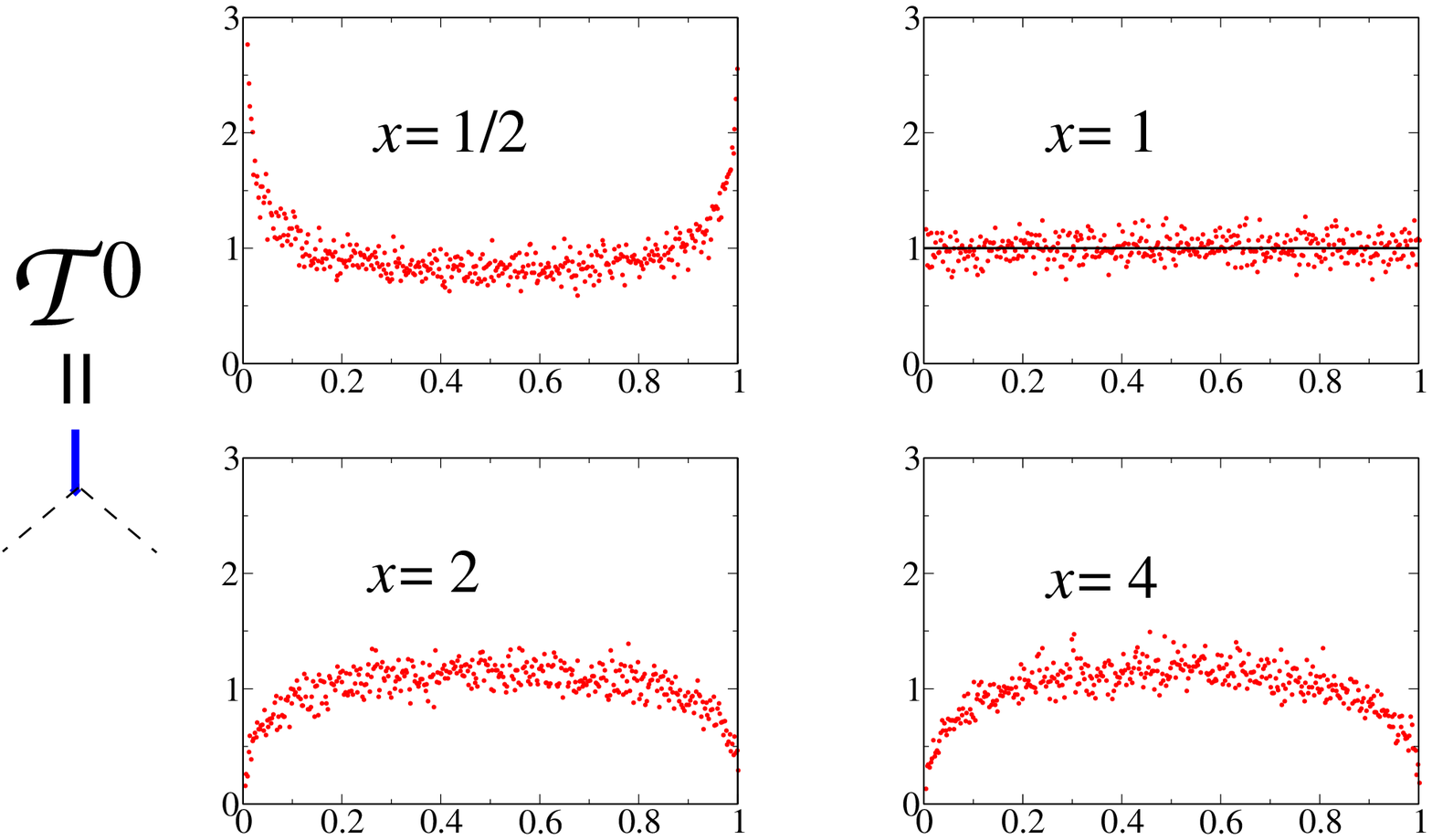}{12.cm}
\figlabel\distpu
\fig{Measured distribution $P(u_L)$ for the initial tree ${\cal T}^0$ drawn on the\nobreak\ left.
}{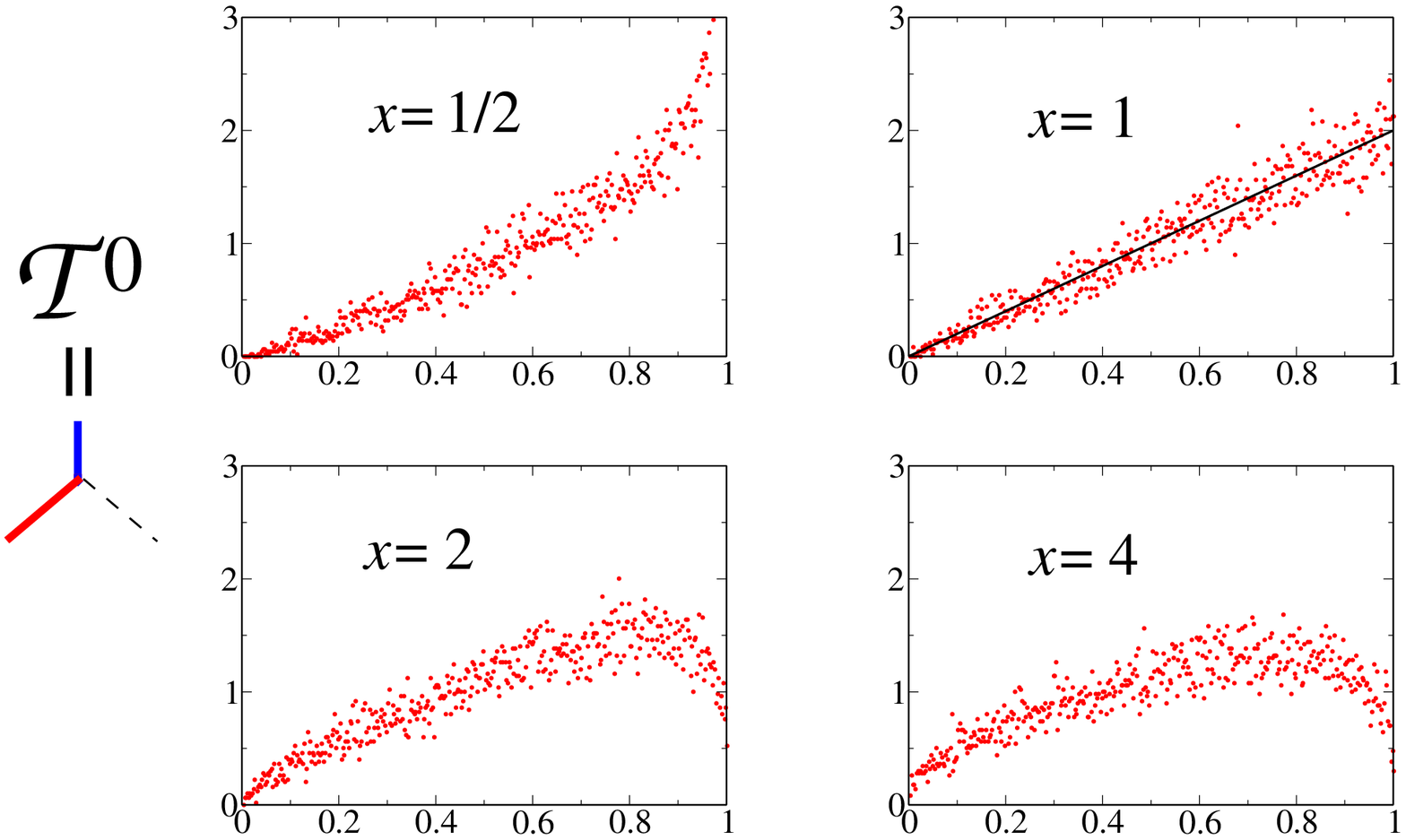}{12.cm}
\figlabel\distputwo
\fig{Measured distribution $P(u_L)$ for the initial tree ${\cal T}^0$ drawn on the\nobreak\ left.
}{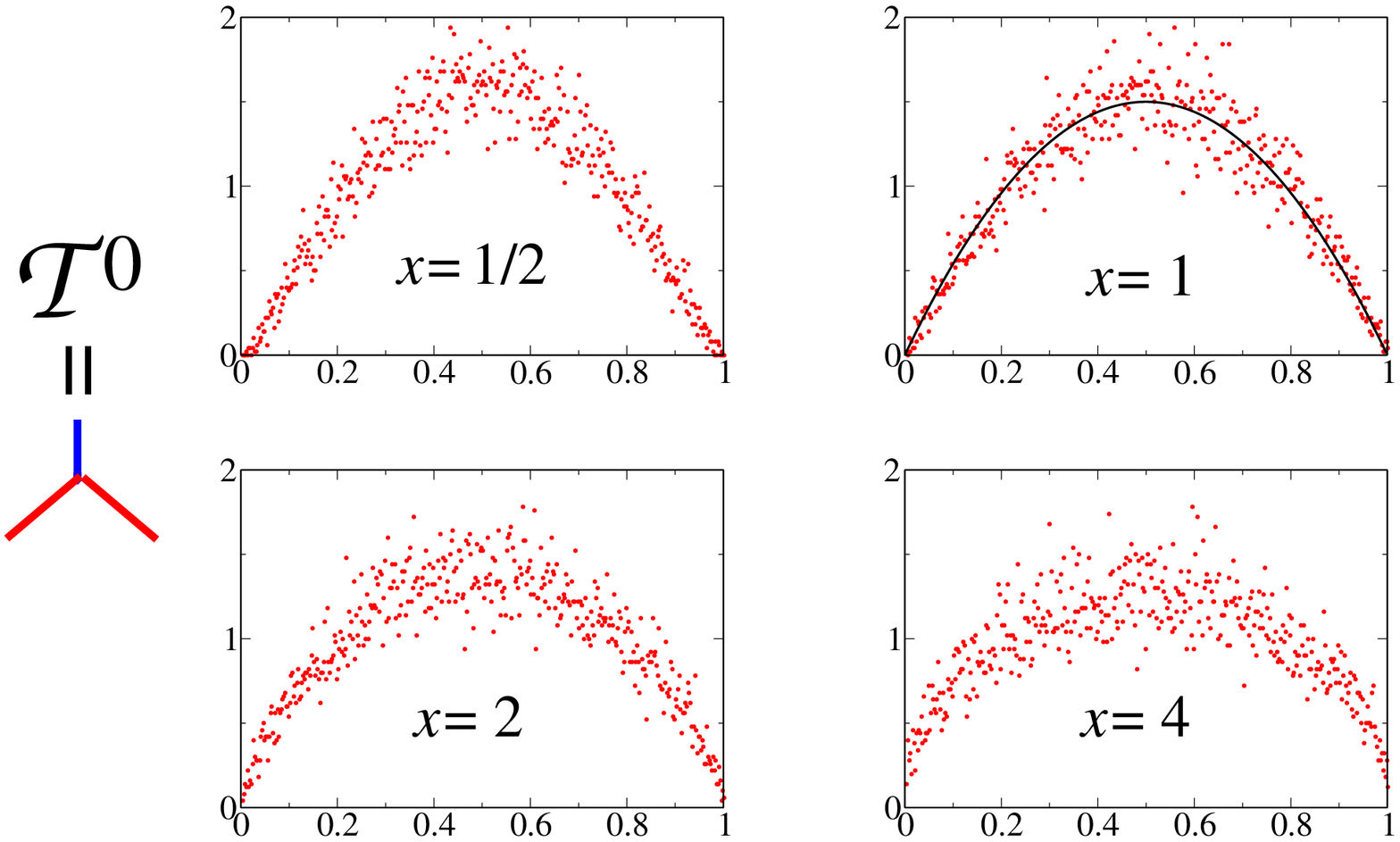}{12.cm}
\figlabel\distputhree
\fig{Measured distribution $P(u_L)$ for the initial tree ${\cal T}^0$ drawn on the\nobreak\ left.
}{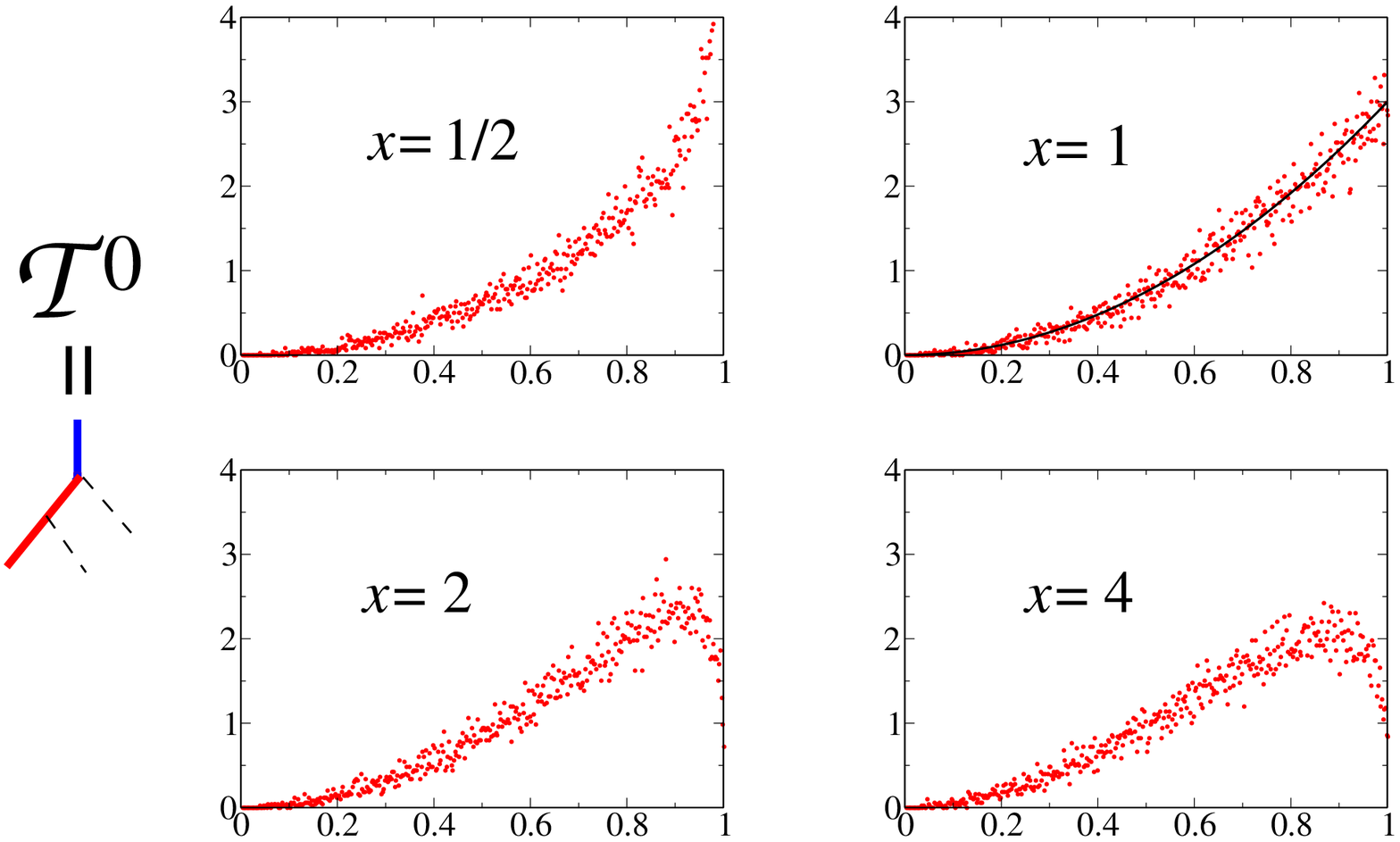}{12.cm}
\figlabel\distpufour
Next, we have explored the left/right mass distribution for various initial conditions
and various of $x$. Again the statistics is over $N=2\times 10^5$ runs and we have chosen
the values $x=1/2,1,2$ and $4$. Figures \distpu-\distpufour\ show
the measured distribution $P(u_L)$ of the proportion $u_L$ of the total mass lying
on the left side. Each point corresponds to an average over 10 consecutive values of the left mass,
with $t/10=400$ such points. The initial configurations are (i) the tree reduced to the root edge 
(figure \distpu), (ii) the tree reduced to root edge and its left descendant (figure \distputwo),
(iii) the tree reduced to root edge and its two descendants (figure \distputhree) and (iv)
the tree reduced to root edge and a chain of length $2$ on the left (figure \distpufour).
For $x=1$, we have also indicated the exact limiting distribution as obtained by
integrating \asympsol\ over $u_R$.
\fig{Measured integrated distributions $D(u)$ for $x=1/2$ (red), $x=1$ (blue), $x=2$ (cyan)
and $x=4$ (green) and for the four initial conditions indicated.}{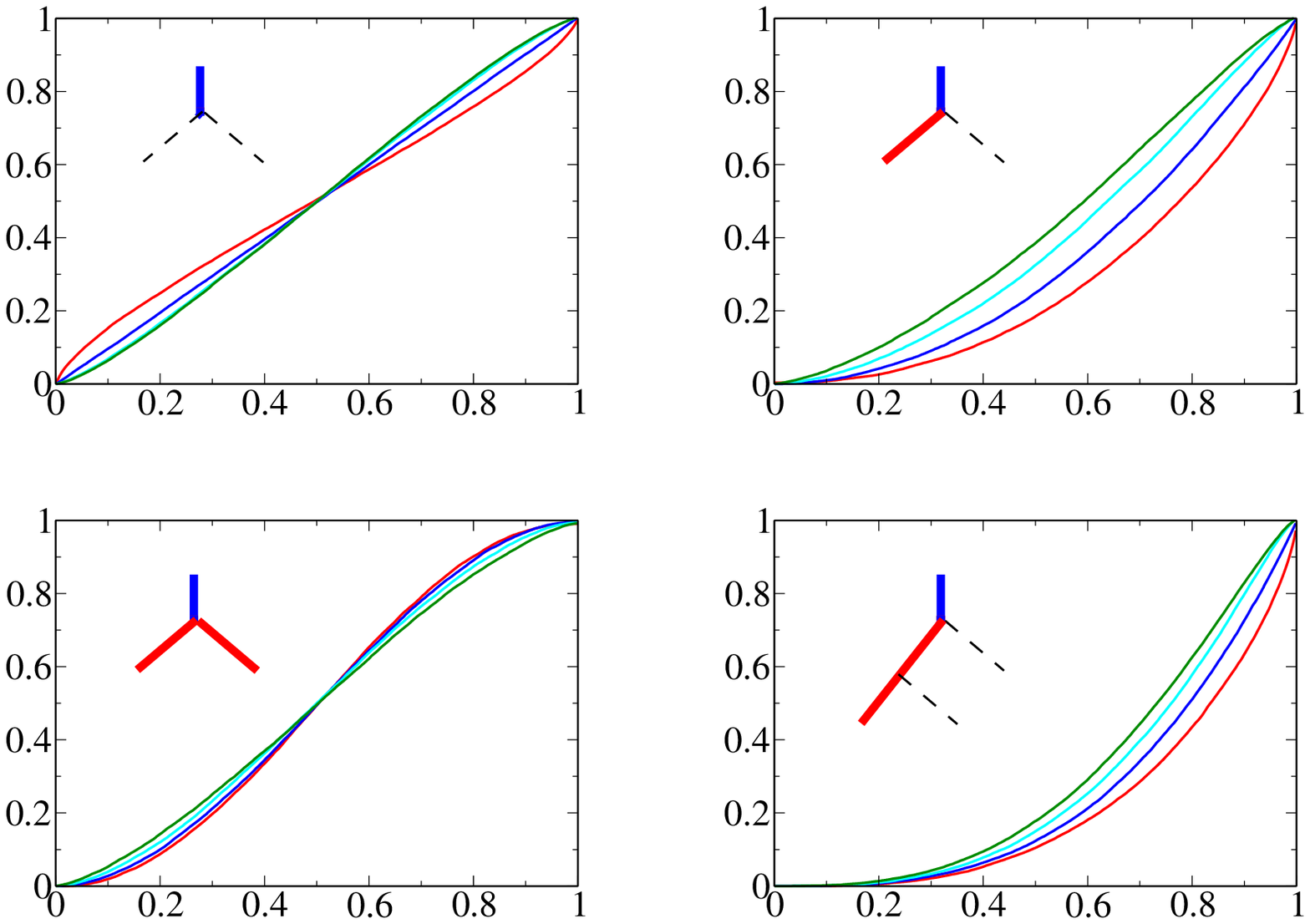}{10.cm}
\figlabel\distduall
The very same data give access to the {\it integrated} distribution
\eqn\integdist{D(u_L)\equiv \int_0^{u_L}P(u)\ du}
by a simple cumulative sum. This wipes out the fluctuations of $P(u)$, leading to smoother curves 
displayed in figure \distduall. In this representation, the measured $x=1$ curves are indistinguishable 
from their exact values deduced from \asympsol.
\fig{Measured exponents $\gamma({\cal T}^0,{\cal T}_L^*)$ at $x=1/2$ from (minus) the slope of
$p({\cal T}^0,{\cal T}_L^*;t)$ vs $t$ in a log-log plot, for an initial tree ${\cal T}^0$ as indicated 
and for various final left subtrees as shown in medallions. The black segment corresponds to the 
predicted exponent, here independent on ${\cal T}_L^*$.
}{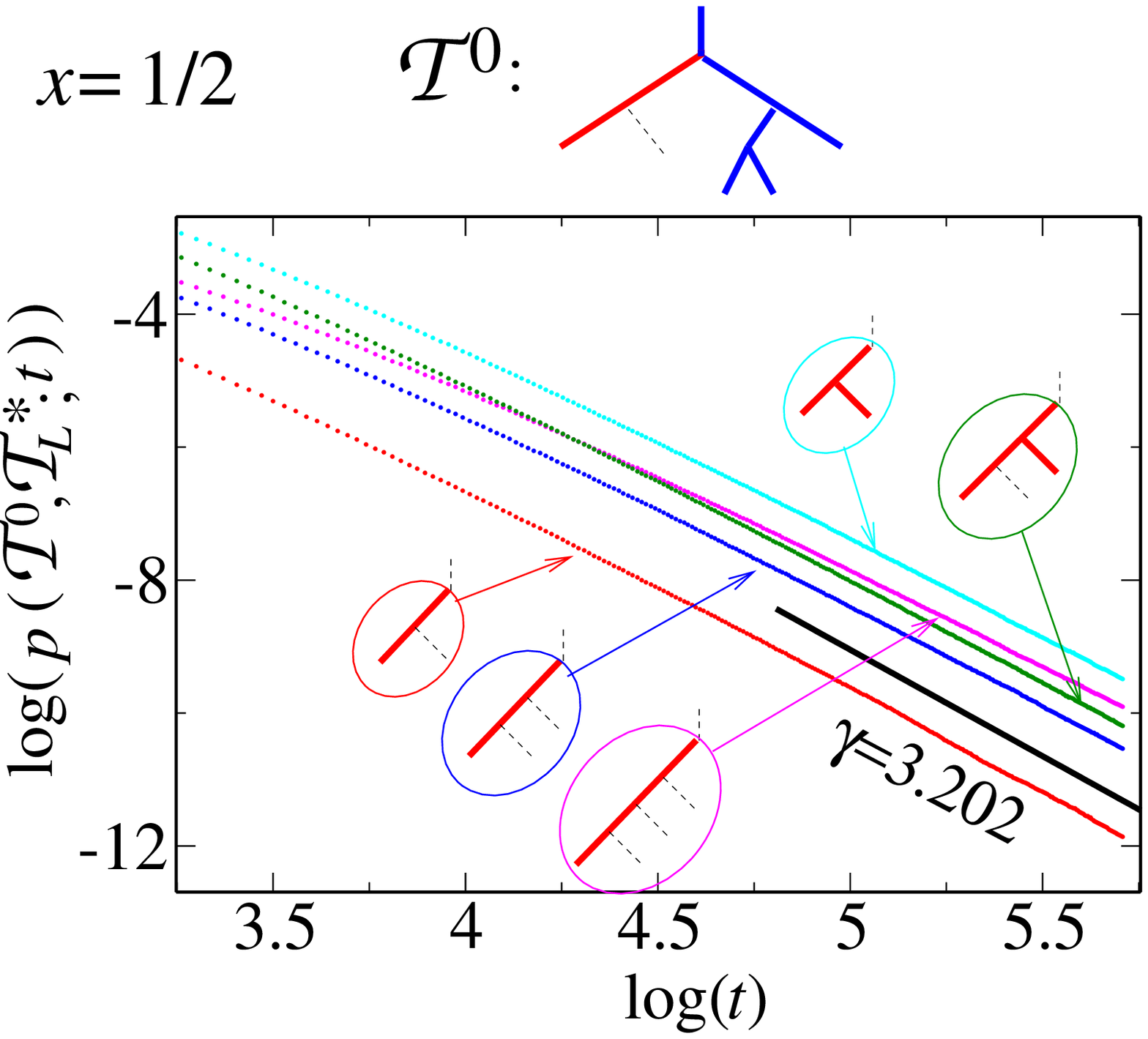}{10.cm}
\figlabel\tdepundemi
\fig{Measured exponents $\gamma({\cal T}^0,{\cal T}_L^*)$ at $x=4$ from (minus) the slope of
$p({\cal T}^0,{\cal T}_L^*;t)$ vs $t$ in a log-log plot, for an initial tree ${\cal T}^0$
as indicated and for various final left subtrees as shown in medallions. The black segment corresponds 
to the two predicted exponents, depending on ${\cal T}_L^*$ (see text).
}{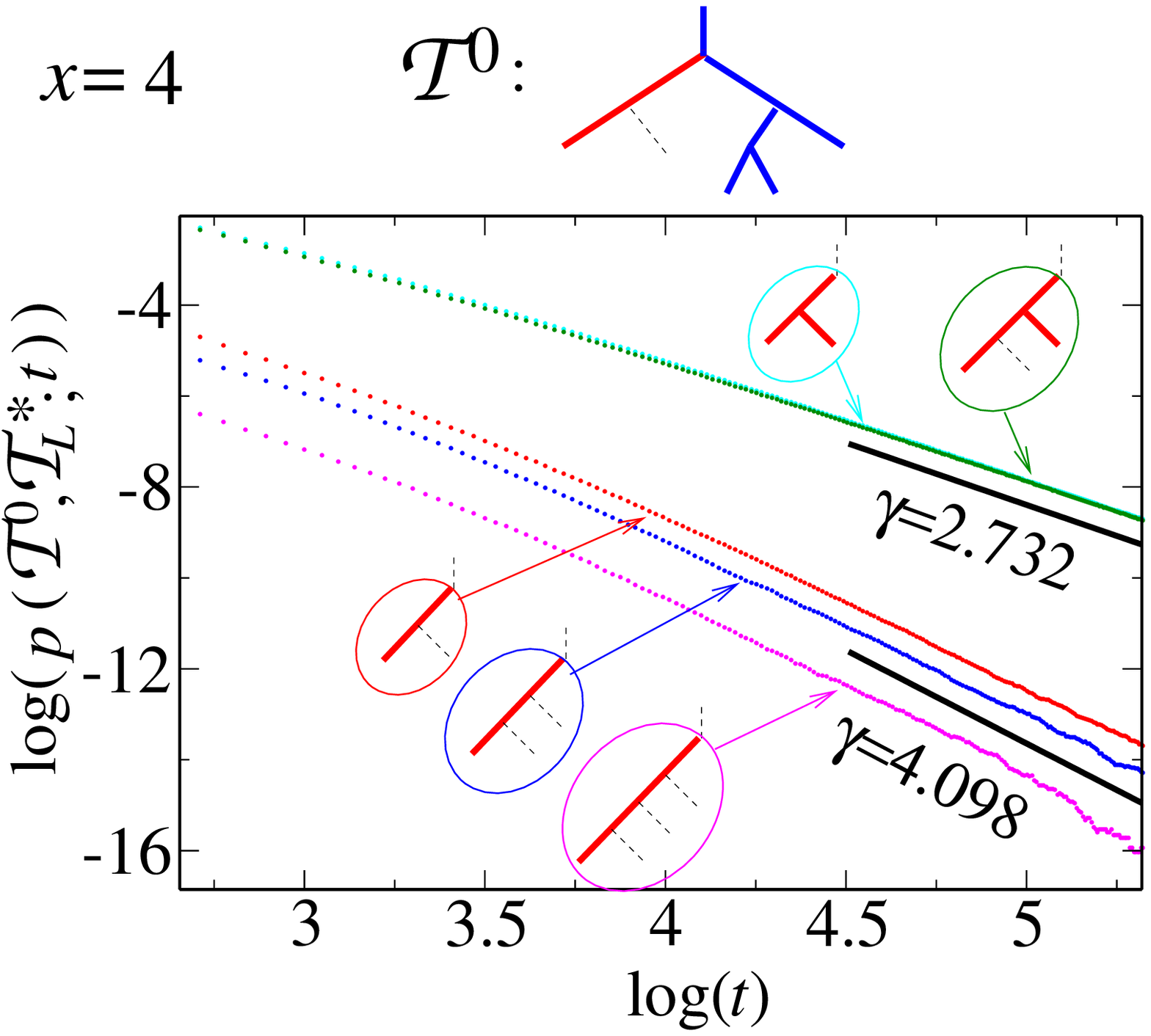}{10.cm}
\figlabel\tdepquatre
We now come to a check of the prediction \finestim\ for the exponent $\gamma({\cal T}^0,{\cal T}_L^*)$
governing the $t$ decay of the probability $p({\cal T}^0,{\cal T}_L^*;t)$ to go in $t$ steps from an
initial tree ${\cal T}^0$ to a large final tree with left part ${\cal T}_L^*$. 
The predicted exponents \finestim\ are plotted against their measured values, i.e. the limiting slopes of
a log-log plot of the probability $p({\cal T}^0,{\cal T}_L^*;t)$ versus $t$. 
The plots \tdepundemi\ and \tdepquatre\ have been obtained from a statistics over $N=10^8$ runs for
$x=1/2$ and $x=4$ and starting with the initial tree ${\cal T}^0$ indicated. 
The various curves correspond to various final left configurations ${\cal T}_L^*$ as indicated in
medallions. 
For $x=1/2$, all curves display the same exponent compatible with the predicted value 
\defgamgener\ for the generic exponent $\gamma({\cal T}^0)$. Indeed, in this case, 
the minimum in \finestim\ is always attained for the initial tree ${\cal T}^0$ 
irrespectively of ${\cal T}_L^*$. This is generic of all values $x<2$.  For $x=4$, we see 
two distinct slopes according to whether ${\cal T}_L^*$ contains or not the left subtree 
${\cal S}_L^0$ obtained by saturating the initial left subtree  ${\cal T}_L^0$ (see section 3.2). 
If it does, we observe the generic value $\gamma({\cal T}^0)$ \defgamgener\ and if not, 
we observe the larger value $\gamma({\cal T}^0,{\cal T}_L^0)$ \finestim. 
\fig{Measured exponent $\beta_L$ from the slope of $P(u_L)$ vs $u_L$ in a log-log plot, for an initial 
tree ${\cal T}^0$ as indicated and for various values of $x$. The straight lines correspond
to the predicted exponents.}{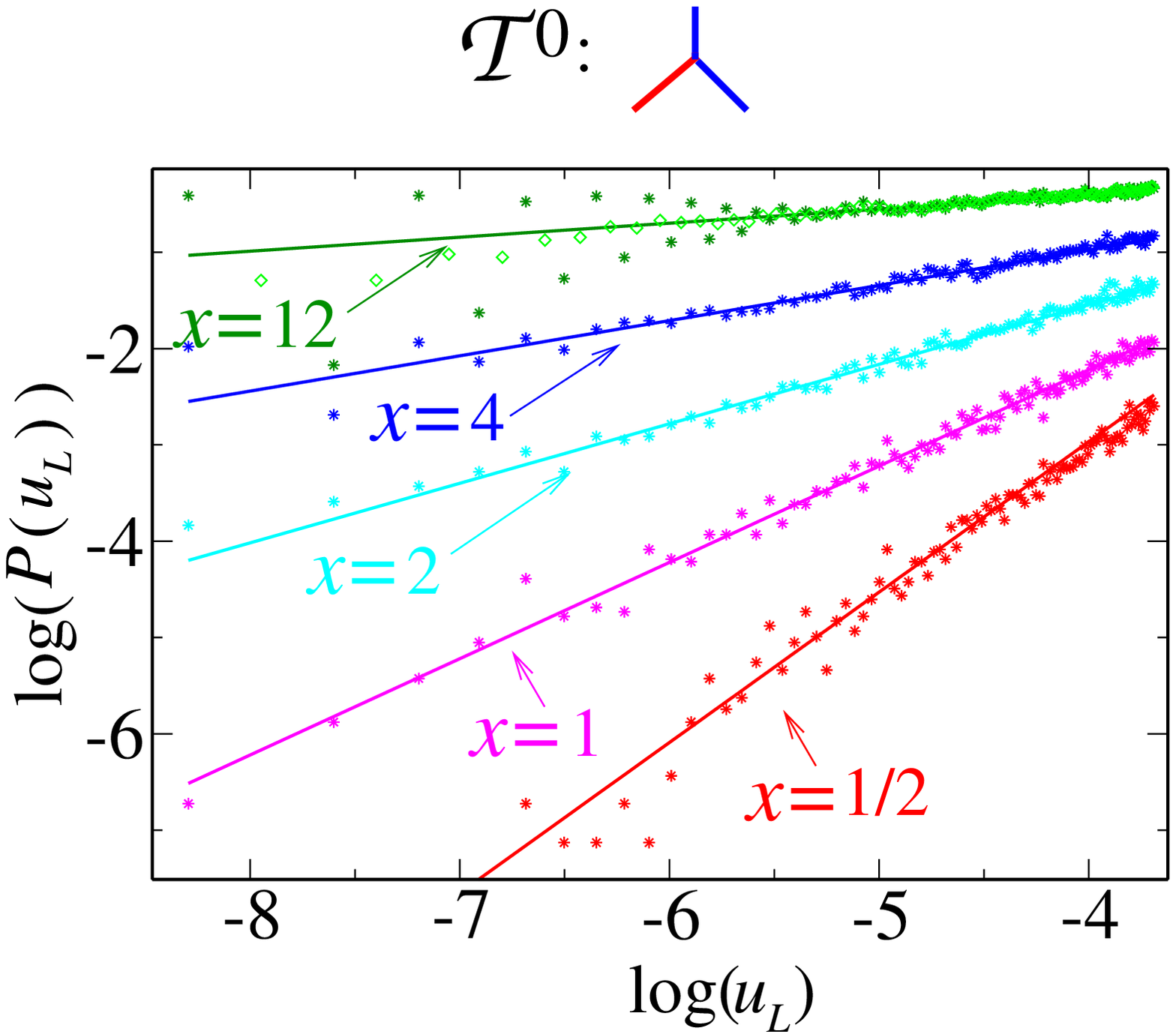}{10.cm}
\figlabel\betacheck
Finally, we have made a direct measurement of the left exponent $\beta_L$ from the 
small $u_L$ behavior \betaLP\ of $P(u_L)$. The statistics is over $N=10^7$ runs for 
an initial tree ${\cal T}^0$ made of the root edge and its two descendants. 
The results are gathered in fig.\betacheck\ in a log-log plot of $P(u_L)$
versus $u_L$ for the values $x=1/2,1,2,4$ and $12$. 
Each point
corresponds to a left mass ranging from $1$ to $100$ over a total mass of $4000$, hence
$u_L$ ranges from $0$ to $.025$. 
The predicted slopes $\beta_L$ 
are indicated by straight lines, with a perfect agreement in the range of large enough masses 
(here larger than $30$), as expected from the scaling argument of section 3.3.
For large $x$, we observe a parity effect due to the fact that the large $x$ preferred value $n_{2,L}=0$ 
(saturated tree) can be attained only for odd left masses. For $x=12$, we have also represented 
the average between two consecutive left masses so as to wipe out this parity effect (empty green
diamonds). This seems to extend the range of validity of the scaling argument to lower $u_L$'s. 
\fig{Log-log plot of the integrated distribution $D(u_L)$ vs $u_L$ from the same
data as fig.\betacheck. The slope are now identified with $1+\beta_L$, with a much better
accuracy.}{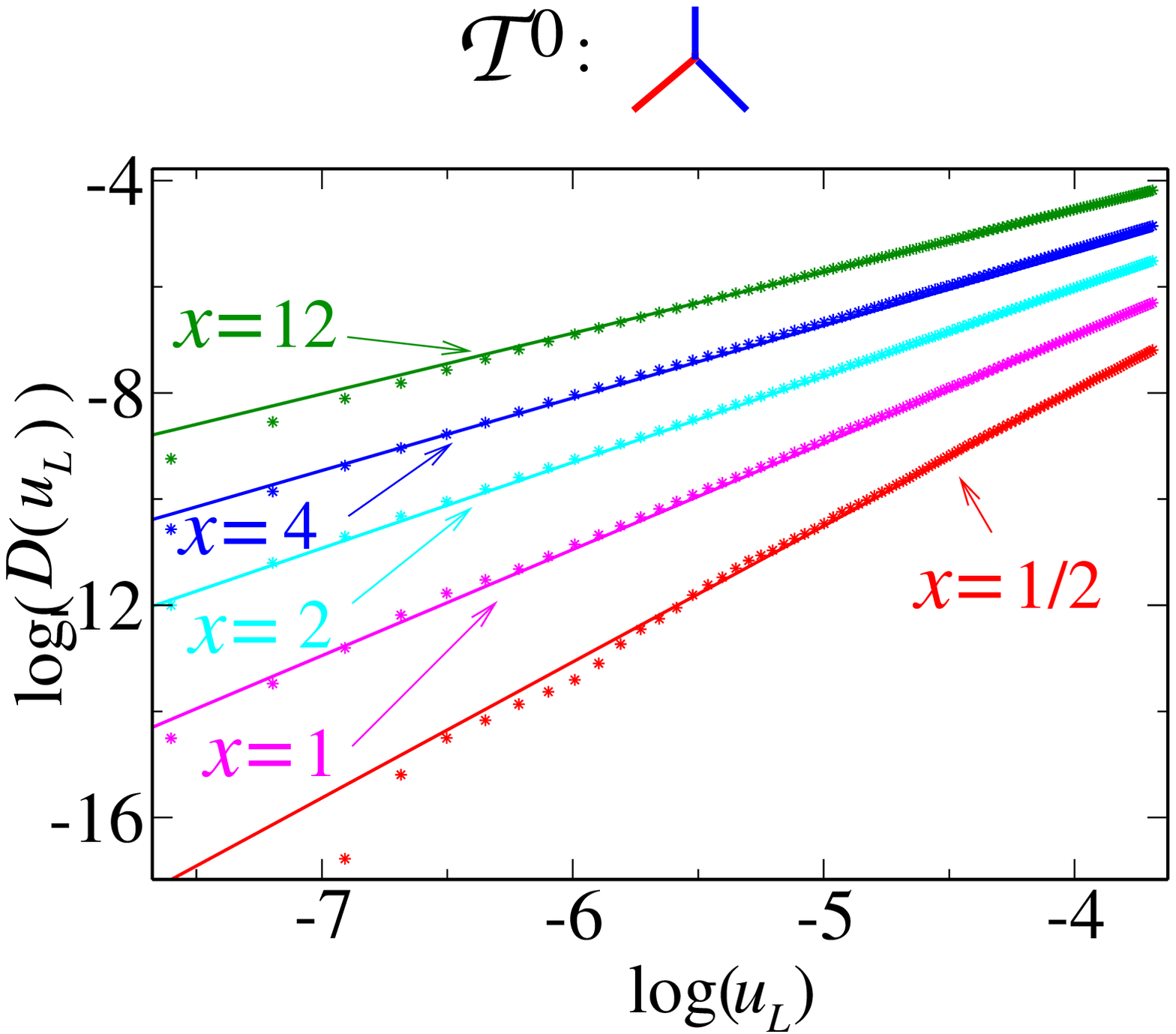}{10.cm}
\figlabel\betainteg
The very same data are used in fig.\betainteg\ to construct the log-log plot of the integrated 
distribution  $D(u_L)$ \integdist\ versus $u_L$ by simple cumulative sums. 
This wipes out the statistical fluctuations of the previous plot and the results corroborate  
unambiguously our predictions for $\beta_L$.

\newsec{Growth of multinary trees}

\subsec{Definition of the model}
In this section, we generalize our results to trees with vertex valences
up to $k+1$ for some fixed integer $k\geq 2$. As before, the trees are planar and rooted and
may be thought of as drawn on top of an underlying $(k+1)$-valent Cayley tree with a unique
leaf attached to the root edge. We start from some initial finite tree ${\cal T}^0$ containing
the root edge and add iteratively links at vertices with valence $i$ less or equal to $k$,
with probability weights $w_i$, $i=1,\ldots,k$. More precisely, if the tree has $n_j$ $j$-valent
vertices ($j=1,\ldots k+1$), a given vertex of valence $i$ is chosen with 
probability $w_i/(\sum\limits_{j=1}^k n_j w_j)$. 
Once the vertex is chosen, the choice of link to add is uniformly distributed
on all the $(k+1-i)$ available descendant edges. Again, this growing process induces a
Markovian evolution for the numbers $n_i(t)$ of $i$-valent vertices at time step $t$
starting from $n_i(0)\equiv n_i^0$, the numbers of $i$-valent vertices on ${\cal T}^0$.
We have the evolution rules
\eqn\multirules{\eqalign{&
{\rm with\ probability\ } {n_1(t) w_1\over \sum\limits_{j=1}^k n_j(t)w_j}\ , \ 
\quad\left\{\matrix{n_1(t+1)&=n_1(t)\hfill\cr n_2(t+1)&= n_2(t)+1\hfill\cr n_\ell(t+1)&=n_\ell(t),\ \ 
\ell\geq 3 \hfill\cr}\right.\cr &
{\rm with\ probability\ } {n_i(t) w_i\over \sum\limits_{j=1}^k n_j(t)w_j}\ , \
\quad\left\{\matrix{n_1(t+1)&=n_1(t)+1\hfill \cr n_i(t+1)&= n_i(t)-1\hfill \cr n_{i+1}(t+1)&=n_{i+1}(t)+1
\hfill\cr n_\ell(t+1)&=n_\ell(t),\ \ \ell\neq 1,i,i+1 \hfill\cr}\right.\cr}}
for $i=2,\ldots k$.

As before, we shall be interested in the repartition of the total mass $T$ (number of edges) of the growing 
tree between the $k$ descending subtrees of the root edge. If we denote by $T_m$ the mass of the
$m$-th descendant subtree from the left (with $T=1+\sum_m T_m$), and $u_m\equiv T_m/(T-1)$, we expect 
a broad limiting distribution $P(u_1,\ldots,u_k)$ at large $T$ characterized by mass distribution
exponents $\beta_m$ such that 
\eqn\betai{P(u_1,\ldots,u_k)\sim u_m^{\beta_m}\quad {\rm when}\quad u_m\to 0}

\subsec{Mean field}
As for binary trees, the expression for the mass distribution exponents $\beta_m$ involves
the limiting proportions $\alpha_i$ of $i$-valent vertices in trees grown for a long time $t$.
Those are again exactly given by a set of mean field equations
\eqn\meanmulti{\eqalign{\alpha_1&= 1-{\alpha_1 w_1\over \Sigma} \cr
\alpha_i &= {\alpha_{i-1} w_{i-1} -\alpha_i w_i \over \Sigma}\, , \quad i=2,\ldots,k\cr }}
where we have defined $\Sigma =\sum\limits_{i=1}^k \alpha_i w_i$. These equations are easily solved into
\eqn\solmeanmulti{\alpha_i= {\Sigma \over w_i} \prod_{j=1}^i {w_j\over \Sigma +w_j}}
where $\Sigma$ is determined by the consistency equation $\Sigma =\sum\limits_{i=1}^k \alpha_i w_i$,
namely
\eqn\constmulti{f(\Sigma)\equiv \sum_{i=1}^k \prod_{j=1}^i {w_j\over \Sigma +w_j} =1}
Note that the function $f(\Sigma)$ is strictly decreasing from $f(0)=k$ to 
$f(\infty)=0$, therefore eq. \constmulti\ has a unique real positive solution and the
equation for the proportions \solmeanmulti\ follow.

\subsec{A solvable case}

Here again, a particularly simple case corresponds to growing the trees by adding links chosen
uniformly at random among all possible available positions. This amounts to taking
$w_i=(k+1-i)$ in which case $\sum w_i n_i=(k-1)T+1$ directly counts the number of available
positions for the addition of a link. In this case, we find $\Sigma=k-1$ and the limiting
proportions $\alpha_i={2k-i-1\choose k-2}/{2k-1\choose k}$. Moreover, we can write a master
equation for the probability $p(T_1,\ldots,T_k;t)$ to have mass $T_m$ for the $m$-th
descendant subtree from the left after $t$ steps:
\eqn\mastermulti{p(T_1,\ldots,T_k;t+1)=\sum_{m=1}^k {(k-1)(T_m-1)+1 \over (k-1)(t+T^0)+1}
\ p(T_1,\ldots,T_{m-1}, T_m-1,T_{m+1},\ldots ,T_k;t)} 
with initial condition $p(T_1,\ldots,T_k;0)=\prod_{m=1}^k \delta_{T_m,T_m^0}$.
Note that this equation may alternatively be interpreted in terms of a generalized RRW in which a walker 
steps in one of $k$ given directions labelled $m=1,2,\ldots k$ with a probability proportional to
a function $f_m(N_m)$ of the number of times $N_m=T_m-T_m^0$ he already stepped in that direction. Here 
the functions read $f_m(N)=N+T_m^0+1/(k-1)$.  In this language, $p(T_1,\ldots,T_k;t)$ is the probability
to have stepped $T_m-T_m^0$ times in the $m$-th direction after $t$ steps (note that in this language
$T_m$ and $T_m^0$ need not be integers but $N_m$ are).
Eq.\mastermulti\ is easily solved into
\eqn\solvmulti{p(T_1,\ldots,T_k;t)={t!\, \Gamma\left(T^0+{1\over k-1}\right)
\over \Gamma\left(T^0+t+{1\over k-1}\right)}\  \prod_{m=1}^k {\Gamma\left(T_m+{1\over k-1}\right)
\over (T_m-T_m^0)!\, \Gamma\left(T_m^0+{1\over k-1}\right)}
\ \delta_{1+\sum\limits_{m} T_m,t+T^0}}
which yields the limiting distribution 
$P(u_1,\ldots,u_k)\equiv \lim\limits_{t\to \infty} t^k p(\{u_m(t+T^0-\nobreak 1)\};t)$:
\eqn\limmulti{P(u_1,\ldots,u_k)={\Gamma\left(T^0+{1\over k-1}\right)\over \prod\limits_{m=1}^k
\Gamma\left(T_m^0+{1\over k-1}\right)}\ \prod_{m=1}^k u_m^{T_m^0-{k-2\over k-1}}\  \delta(\sum_{m=1}^k
u_m-1)}
We can therefore read off the mass distribution exponents for this particular case:
\eqn\betasolmu{\beta_m=T_m^0-{k-2\over k-1}}

\subsec{Mass distribution exponents}
In the case of arbitrary weights $w_j$, the mass distribution exponents may be evaluated 
along the same lines as in Section 3, with the result
\eqn\multigenexp{\beta_m= -1+{1 \over \Sigma}
\ \min\limits_{{\cal T}\supset {\cal T}_m^0} \{\sum_{j=1}^k w_j n_j({\cal T})\} 
}
with $\Sigma$ the solution of eq. \constmulti\ and where the minimum is taken
over all trees ${\cal T}$ containing the $m$-th initial subtree ${\cal T}_m^0$.
Again we use the convention that whenever ${\cal T}$ is empty, we set 
$n_1({\cal T})=\cdots=n_{k-1}({\cal T})=0$ and $n_k({\cal T})=1$.

Note that, in the solvable case of previous section, the quantity to be minimized
is simply $(k-1)T+1$ where $T$ is the mass of ${\cal T}$ hence the minimum is
reached for ${\cal T}_m^0$, leading to the result $-1+((k-1)T_m^0+1)/\Sigma$, in
agreement with eq. \betasolmu\ as $\Sigma=k-1$.

\fig{The four regimes (i-iv) for the determination of the mass distribution exponent $\beta_m$ in the 
$(y=w_2/w_1,z=w_3/w_1)$ plane.  The exponent $\beta_m$ is plotted here for the particular choice 
of ${\cal T}_m^0$ indicated in the upper right corner.  We may always write 
$\beta_m=-1+(n_1 w_1+n_2 w_2 +n_3 w_3)/\Sigma$ provided we choose for $n_i$ the number of 
$i$-valent vertices of the appropriate tree, namely: (i) the initial 
tree ${\cal T}_m^0$, or (ii) a tree obtained 
from ${\cal T}_m^0$ by changing its bivalent vertices into trivalent ones,
or (iii) the tree obtained from ${\cal T}_m^0$ by changing its trivalent vertices into tetravalent ones,
or (iv) the tree obtained form ${\cal T}_m^0$ by changing both its bi- and trivalent vertices into 
tetravalent ones. These trees are displayed on the right. 
}{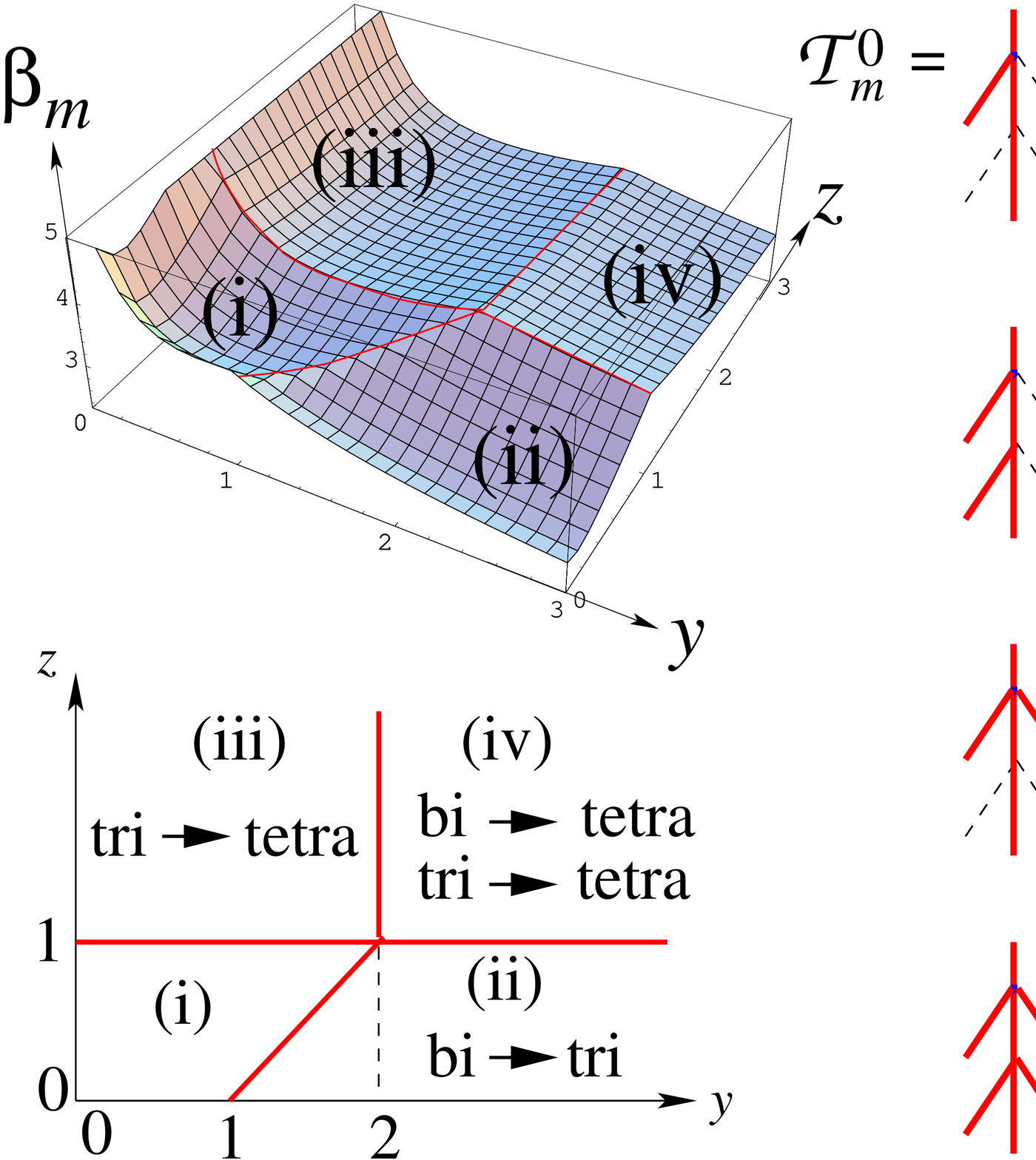}{11.cm}
\figlabel\phases
For illustration, let us discuss the case $k=3$ in detail. It is convenient to
view the trees ${\cal T}$ as grown out of ${\cal T}_m^0$ and to follow the evolution
of the quantity $r({\cal T})\equiv \sum_{j=1}^3 w_j n_j({\cal T})$ that we wish to minimize.
According to the evolution rules \multirules\ applied to up to three consecutive steps,
$r({\cal T})$ has increments 
\eqn\increments{\eqalign{
w_2 \hfill & \quad \hfill \hbox{when attaching one link to a univalent vertex}\cr
w_3+w_1 \hfill & \quad \hfill \hbox{when attaching two links to a univalent vertex}\cr
2w_1 \hfill & \quad \hfill \hbox{when attaching three links to a univalent vertex}\cr
w_3+w_1-w_2 \hfill & \quad \hfill \hbox{when attaching one link to a bivalent vertex}\cr
2w_1-w_2 \hfill & \quad \hfill \hbox{when attaching two links to a bivalent vertex}\cr
w_1-w_3 \hfill & \quad \hfill \hbox{when attaching one link to a trivalent vertex}\cr
}}
In particular, the increment is always positive whenever we attach new links to
an existing univalent vertex. Let us introduce the ratios
\eqn\ratios{y\equiv {w_2\over w_1}\ ,\qquad z\equiv {w_3\over w_1}}
For $z<1$ and $y<z+1$ (regime (i)), all increments above are positive. The minimum is therefore
attained for the initial configuration ${\cal T}={\cal T}_m^0$, with $n_i({\cal T})=n_i^m$ the
initial numbers of $i$-valent vertices in ${\cal T}_m^0$. For $z<1$ and $y>z+1$
(regime (ii)), the quantity
$r({\cal T})$ is minimized by adding exactly one link to all bivalent vertices of ${\cal T}_m^0$, 
thus changing them into trivalent ones. The resulting tree ${\cal T}$  now has 
$n_1({\cal T})=n_1^m+n_2^m$, $n_2({\cal T})=0$ and $n_3({\cal T})=n_2^m+n_3^m$.
For $z>1$, it becomes favorable to add one link to all trivalent vertices,
thus changing them all into tetravalent vertices. As for bivalent vertices, either
$y<2$ (and in particular $y<z+1$) and we leave these vertices unchanged (regime (iii)), resulting in a 
tree with $n_1({\cal T})=n_1^m+n_3^m$, $n_2({\cal T})=n_2^m$ and $n_3({\cal T})=0$,
or $y>2$ and we must change these bivalent vertices into tetravalent ones (regime (iv)), resulting
in a tree with $n_1({\cal T})=n_1^m+2 n_2^m+n_3^m$, $n_2({\cal T})=n_3({\cal T})=0$.
This allows us to write
\eqn\multivalbet{\eqalign{
{\rm (i)} \ \ \beta_m  
&= {n_1^m w_1 +n_2^m w_2 +n_3^m w_3 \over \Sigma}-1 \quad {\rm for} \quad z<1,\ y<z+1 \cr
{\rm (ii)} \ \ \beta_m  
&= {(n_1^m +n_2^m)w_1 +(n_2^m+n_3^m) w_3 \over \Sigma}-1 \quad {\rm for} \quad z<1,\ y>z+1 \cr
{\rm (iii)}\ \ \beta_m  &= {(n_1^m +n_3^m) w_1 +n_2^m w_2 \over \Sigma}-1 \quad {\rm for} \quad z>1,\ y<2 \cr
{\rm (iv)}\ \ \beta_m  &= {(n_1^m +2n_2^m +n_3^m) w_1 \over \Sigma}-1 \quad {\rm for} \quad z>1,\ y>2\cr}}
These formulae also hold whenever ${\cal T}_m^0$ is empty by setting $n_1^m=n_2^m=0$ and 
$n_3^m=1$.
These results are summarized in figure \phases\ where the different regimes are displayed.
We have also performed a number of numerical checks identical to 
those presented in section 3 and which fully corroborate the above
results.

\newsec{Conclusion}

In this paper, we have derived the exact expression of exponents 
characterizing the mass distribution of trees growing locally by
addition of links. As opposed to generic random trees whose statistics
is governed by local Boltzmann weights, the mass of large growing trees 
is not in general concentrated in a single side of the root but can be
distributed in all subtrees. The mass can be preferentially
concentrated in one subtree (negative mass distribution exponent), 
it can be preferentially equally distributed over all subtrees (positive
exponent) or it can be characterized by a uniform distribution (vanishing
exponent).
In practice, the precise value of the mass distribution exponent
depends both on the growth process parameters $w_i$ and on the initial
condition ${\cal T}^0$.
In particular, in the case of binary trees grown from a single edge, 
the exponent $\beta_L$
increases with $x$ up to the transition point $x=2$ and then
decreases down to $0$ at $x\to \infty$. This shows that
a monotonous variation of the relative strength that we
attach to the two types of vertices (univalent or bivalent) does not induce 
a monotonous variation of the characteristics of the tree.

We have observed that, except for special values of $x$ where the model is
exactly solvable, the mass distribution is not in general a Beta law and we
may wonder whether an explicit analytic expression could be found. In this
respect, a promising direction of investigation is provided by the approach
of the problem via continuous time branching processes [\xref\janson-\xref\rudas].
Indeed, within this framework, it is possible to relate the mass distribution to
other asymptotic large time distributions which are themeselves determined 
by coupled integral equations. Solving these equations analytically would
be a major advance in this field.

For more general tree valences, an interesting outcome of our study is the 
existence of different regimes for the value of the mass distribution exponents. 
This seems to indicate the existence of a competition between various 
coexisting effective growth mechanisms. Whether
the transitions between these regimes are characterized or not by 
the emergence of some order parameter remains to be understood.

\bigskip
\noindent{\bf Acknowledgments:} 
We thank J. Bouttier for helpful discussions. 
All authors acknowledge support from the Programme d'Action
Int\'egr\'ee J. Verne ``Physical applications of random graph theory"
and from the ENRAGE European network, MRTN-CT-2004-5616. 
We acknowledge support from the Geocomp project,
ACI Masse de donn\'ees (P.D.F and E.G.), 
from the ENIGMA European network, MRTN-CT-2004-5652 (P.D.F) and from 
the ANR program GIMP, ANR-05-BLAN-0029-01 (F.D. and P.D.F.).
\listrefs
\end
\bye